\documentclass[useAMS,usenatbib]{mn2e}

\usepackage{graphicx,color}
\newcommand{\etal }{{et al.} }
\newcommand{\msun}{\thinspace M_\odot} 
 
\newcommand{\lsun}{\thinspace L_\odot}

\def\lesssim{\mathrel{\hbox{\rlap{\hbox{\lower4pt\hbox{$\sim$}}}\hbox{$<$}}}}
\def\gtrsim{\mathrel{\hbox{\rlap{\hbox{\lower4pt\hbox{$\sim$}}}\hbox{$>$}}}}
\newcommand{\cm}{\,{\rm cm}^{-3} } 
 
\newcommand{\nc}{n_{\rm c} } 
\newcommand{\rcri}{R_{\rm cl} }
\newcommand{\mdot}{M_\odot\,{\rm yr}^{-1} }

\newcommand{\dfrac}[2]{{\displaystyle \frac{#1}{#2}} }

\newcommand{\mv}{M_\odot\, {\rm km}\, {\rm s}^{-1}}

\title[Evolution of Protostellar Outflow]{Evolution of Protostellar Outflow around Low-mass Protostar}
\author[M. N. ~Machida,  \& T. Hosokawa]
  { Masahiro N. Machida$^{1}$ \thanks{E-mail: machida.masahiro.018@m.kyushu-u.ac.jp (MNM)} and Takashi Hosokawa$^{2,3}$\\
$^{1}$ Department of Earth and Planetary Sciences, Faculty of Sciences, Kyushu University, Fukuoka 812-8581, Japan\\
$^{2}$ Department of Physics, University of Tokyo, Tokyo 113-0033, Japan\\
$^{3}$ Jet Propulsion Laboratory, California Institute of Technology,
Pasadena, CA 91109, USA 
}

\begin{document}

\maketitle

\begin{abstract}
The evolution of protostellar outflow is investigated with resistive
magneto-hydrodynamic nested-grid simulations that cover a wide range of
spatial scales ($\sim1$\,AU\,--\,1\,pc).
We follow cloud evolution from the pre-stellar core stage until
the infalling envelope dissipates long after the protostar formation.
We also calculate protostellar evolution to derive protostellar 
luminosity with time-dependent mass accretion through a circumstellar
disk. 
The protostellar outflow is driven by the first core prior to 
protostar formation and is directly driven by the circumstellar disk 
after protostar formation. 
The opening angle of the outflow is large in the Class 0 stage.
A large fraction of the cloud mass is ejected in this stage,
which reduces the star formation efficiency to $\sim 50$~\%.
After the outflow breaks out from the natal cloud,
the outflow collimation is gradually improved in the Class I stage.
The head of the outflow travels more than $\sim10^5$\,AU in $\sim10^5$\,yr.
The outflow momentum, energy and mass derived 
in our calculations agree well with observations.
In addition, our simulations show the same correlations among outflow
momentum flux, protostellar luminosity and envelope mass 
as those in observations.
These correlations differ between Class 0 and I stages, which is
explained by different evolutionary stages of the outflow;
in the Class 0 stage, the outflow is powered by the accreting mass 
and acquires its momentum from the infalling envelope; 
in the Class I stage, the outflow enters
the momentum-driven snow-plough phase.
Our results suggest that protostellar outflow should determine
the final stellar mass and significantly affect 
the early evolution of low-mass protostars.
\end{abstract}

\section{Introduction}
\label{sec:intro}
Molecular outflows are ubiquitously observed in the star forming region,
which indicates that young protostars generally drive the outflows.
The molecular outflow can dump a large fraction of cloud matter into the interstellar space, and only the remaining gas around the protostar contributes to protostellar mass growth.
Therefore, the molecular outflow controls the resulting stellar mass and significantly affects the star formation process.
The star forms in a gravitationally contracting cloud.
Although the specific outflow driving mechanism is uncertain, the molecular outflow, in principal, is powered by the gravitational energy of the infalling matter released in the gravitationally contracting cloud.
The infalling matter, or infalling envelope, exists only in the early phase of star formation (Class 0 and I stages; \citealt{andre93,andre94}) during which powerful outflows are often observed.
Thus, observation of the molecular outflows provides a clue for understanding the early phase of star formation.


Since the first discovery of molecular outflow \citep{snell80}, more than 300
outflows have been observed in various star forming regions
\citep{wu04,hatchell07}.
\citet{cabrit92} found that the outflow 
momentum flux with 16 outflow samples systematically increases with the 
stellar bolometric luminosity. 
\citet{bontemps96} showed that the outflow momentum flux also 
correlates well with the (infalling) envelope mass for Class 0 and I stages.
Moreover, they also argued that the outflow power decreases with time during the 
accretion phase and that the outflow properties qualitatively differ between Class 0 and I protostars.
However, with considerable data scatter,  \citet{hatchell07} 
failed to confirm that Class I protostars generally have a 
lower momentum flux than Class 0 sources.
Recently, \citet{curtis10} analysed the outflow properties of 45
samples and reported a decrease in outflow momentum flux 
from the Class 0 to I stage, as shown in \citet{bontemps96}.


Powerful outflows are frequently observed near the youngest 
(Class 0) objects \citep{bachiller92}, indicating that vigorous 
outflow emerges in the very early evolutionary phase in which 
the protostar has attained a small fraction of its final mass 
\citep{bontemps96}.
These observations suggest that  molecular outflow affects 
the early evolution of newly born stars.
However, we cannot directly observe the outflow driving region, 
which is deeply embedded in a dense infalling envelope.
Thus, it is difficult to specify the outflow driving mechanism 
with only observational results.
Theoretical modelling or numerical simulations are necessary 
to understand and thereby resolve this issue.


Since the discovery of well-collimated jet-like flows (optical jets: \citealt{mundt83}), it has been postulated that low-velocity 
wide-angle flows, or molecular outflows, are entrained  by these jets.
This simple notion has been prevalent because both
high-velocity jets and low-velocity outflows are comprehensively
explained.
Many authors have proposed various entrainment models 
\citep{cabrit97,richer00,lee00,arce07} to analytically or numerically study the outflow driving mechanism in which the jet is artificially injected into the ambient medium to entrain the infalling material.
However, it is difficult to specify the outflow driving mechanism
because an abundance of free or unknown parameters is available to reproduce low-velocity outflow entrained by high-velocity jets (\S\ref{sec:outflowmodel}).


Conversely,  a completely different theoretical concept of the molecular outflow has been proposed.
\citet{tomisaka02} calculated the gravitational collapse of a molecular
cloud core and demonstrated that both  high-velocity and low-velocity flows 
are launched from different objects; the low-velocity flow is launched 
from the first core \citep{larson69,masunaga00}, and  the
high-velocity flow is launched near the protostar.
This concept has been supported by many other cloud-collapse simulations 
\citep[see review of ][]{machida11d}.
Therefore, in such works, the authors followed the evolution 
with the absence of free parameters to control the jet and outflow.
However, they could not calculate long-term evolution of outflow 
because the protostar itself was resolved \citep{tomisaka02,machida08b}, and the numerical timestep became increasingly  short as the protostar and jet evolved \citep[e.g.][]{machida08b}.


Long-term cloud collapse simulations were conducted in our previous work
 \citep{machida12} using the sink-cell technique.
We demonstrated that the first core evolves to the circumstellar disk 
after protostar formation
\citep{bate98,bate10,machida10a}.
The low-velocity flow, which is launched from the first core prior to the
protostar formation, is driven by the circumstellar disk 
after its formation. 
Very recently,  outflow driven by the first core (candidate) were observed by several authors \citep{dunham11,chen10,enoch10,pineda11,chen12}.
A considerably young outflow was also observed around an extremely young prestellar core \citep{takahashi12a,takahashi12b}.
These observations agree well with our simulation results.

In this paper, we extend our previous work to directly compare outflows simulated in the later evolutionary stages with observations by calculating cloud evolution from the prestellar
cloud stage until the protostar evolves into the Class I or II stage.
We compare the resulting outflow properties such as the
momentum flux, energy and shape with observation data. 
In addition, the protostellar evolution is numerically calculated with
time-dependent accretion histories obtained in the simulations. 
This process enables the examination of the correlations among outflow momentum flux, protostellar luminosity and
envelope mass, which are suggested by observations 
\citep[e.g.][]{bontemps96, cabrit92}.


This paper is structured in the following manner. 
The framework of our models and the numerical method are described in \S 2, and the numerical results are presented in \S 3. 
We compare calculation results with observations in \S 4, discuss the model parameters and outflow models in \S 5 and summarize our results in \S 6.

\section{Numerical Method}
\label{sec:model}
In this study, we calculate the evolution of both cloud cores and
protostars.
With our magnetohydrodynamic (MHD) code, we calculate the evolution of
a collapsing core from the prestellar stage until the point at which the collapsing core, or the infalling envelope, 
dissipates after the protostar formation.
In our simulations, we demonstrate that protostellar outflows are driven by circumstellar disks. 
To follow the long-term evolution over the entire main accretion phase,
we adopt sink cells for masking the specific vicinity of the protostar. 
In addition, we calculate the protostellar evolution 
with time-dependent accretion histories obtained in the MHD simulations.
This combination of  MHD simulations and stellar evolution
calculations enable us to link
the evolution of protostellar outflow with protostellar evolution.
In this section, we first describe the method and settings of 
the MHD calculations and explain our modelling process of  protostellar evolution.

\subsection{MHD Calculation}
\subsubsection{Initial Settings}
\label{sec:setting}
To investigate the evolution of cloud cores, we use three-dimensional resistive MHD equations including self-gravity.
The numerical method is described in \citet{machida04}, \citet{machida05a} and \citet{machida05b}.
The basic equations, resistivity and sink cell treatment are the same as those reported in \citet{machida12}.
In the calculation, instead of solving the energy equation, we use a barotropic equation of state (eq. [5] of \citealt{machida12}) which mimics the thermal evolution of the collapsing prestellar cloud core \citep{larson69,masunaga00}. 
This treatment makes a long-term calculation possible.
It should be noted that, however, the barotropic equation of state is not adequate especially long after the protostar formation  because the protostellar luminosity can heat up the surrounding gas \citep{masunaga00,whitehouse06,krumholz06}.

As the initial state, we assume an isolated cloud core embedded in an interstellar medium.
We adopt a spherical cloud with a critical Bonnor--Ebert (BE) density profile, $\rho_{\rm BE}$, in which a uniform density is adopted outside the sphere ($r > \rcri$, where $\rcri$ is the critical BE radius) to mimic the interstellar medium.
We prohibit gas inflow at $r=R_c$ to strictly avoid mass inflow from outside the core, while we do not prohibit mass outflow at the boundary between the BE sphere and  interstellar medium.
Thus, the outflowing gas can freely escape from the BE sphere through protostellar outflow.
Hereafter, we refer to the gravitationally bound gas cloud within $r<R_c$ as the host cloud. 
It should be noted that we confirmed that the total mass of the host cloud is well conserved during the calculation before the protostellar outflow reaches the cloud boundary.

Because the critical BE sphere is in equilibrium, we increase the density by a factor of $f=1.68$ to promote contraction, where $f$ is the density enhancement factor that represents the stability of the initial cloud.
The cloud stability is generally represented by a parameter $\alpha_0$ ($\equiv E_{\rm t}/E_{\rm g}$), which is the ratio of thermal energy ($E_{\rm t}$) to gravitational energy ($E_{\rm g}$).
The density enhancement factor of $f=1.68$ corresponds to $\alpha_0=0.5$ (\citealt{machida06,machida12}).
The density profile of the initial cloud is the same as that reported in \citet{machida12}.

For a dimensional BE density profile, we adopt an isothermal temperature of $T=10$\,K and a central number density of $\nc  = 6\times10^5\,\cm$.
With these parameters, the critical BE radius ($r_{\rm BE}$) is  $r_{\rm BE} = 6.1\times10^3$\,AU.
For typical models, the cloud core has a critical BE radius of $R_{\rm cl}=r_{\rm BE}$, where $R_{\rm cl}$ is the cloud core radius.
The mass inside $r < R_{\rm cl}$ for the typical models is  $M_{\rm cl}=1.05\msun$.
In addition, to investigate the mass dependence of the host cloud on the evolution of protostellar outflow, we create two exceptional models with different initial cloud mass, which changes the cloud radius $R_{\rm c}$.
These models are 1.5 and 2 times  the critical BE radius; the cloud radius for these models is  $R_{\rm c} = 1.5\,r_{\rm BE}$ (= $9.2\times10^3$\,AU) and $R_{\rm c} = 2\, r_{\rm BE}$ (=$1.2\times10^4$\,AU), respectively.
The initial cloud mass for these models is  $M_{\rm cl} = 1.6\,\msun$ for the model with $r=1.5\,R_{\rm cl}$  and $M_{\rm cl} = 2.1\,\msun$ for that with $r=2\,R_{\rm cl}$, respectively.
The model name, host cloud radius and mass are listed in Table~\ref{table:1}.

In each model, the cloud rotates rigidly around the $z$-axis in the  $r< \rcri$ region and a uniform magnetic field parallel to the $z$-axis, or rotation axis, is adopted in the entire computational domain.
We parameterized the initial magnetic field strength and rotation rate.
The magnetic field strength is scaled using the central density $\rho_0$ and sound speed $c_{s,0}$ as
\begin{equation}
b =  B_0^2 / (4\pi \, \rho_0 \, c_{s,0}^2),
\label{eq:alpha}
\end{equation}
and the rotation rate is scaled using the central density as
\begin{equation}
\omega = \Omega_0/(4 \pi\,  G \, \rho_0  )^{1/2}.
\label{eq:omega}
\end{equation}
With these parameters, we created ten models as listed in Table~\ref{table:1}.
The dimensional magnetic field strength ($B_0$) and angular velocity ($\Omega_0$) for each model are also described in Table~\ref{table:1}.
To simply characterize models, we estimated the ratios of rotational and magnetic energies to the gravitational energy, $\beta_0$ ($\equiv E_{\rm rot}/E_{\rm grav}$) and $\gamma_0$ ($\equiv E_{\rm mag}/E_{\rm grav}$), where $E_{\rm rot}$ and $E_{\rm mag}$ are rotational and magnetic energies, respectively, and summarized them in Table~\ref{table:1}.
We also estimated the mass-to-flux ratio $M/\Phi$ of each host cloud.
\citet{mouschovias76} derived the following critical mass-to-flux ratio
\begin{equation}
\left(\dfrac{M}{\Phi}\right)_{\rm cri} = \dfrac{\zeta}{3\pi}\left(\dfrac{5}{G}\right)^{1/2},
\label{eq:mag2}
\end{equation}
where the constant $\zeta=0.48$ \citep{tomisaka88a,tomisaka88b}.
The mass-to-flux ratio normalized by the critical value $\mu$ is described as
\begin{equation}
\mu  \equiv \left(\dfrac{M}{\Phi}\right) \left(\dfrac{M}{\Phi}\right)_{\rm cri}^{-1}.
\label{eq:crit}
\end{equation}
Models include a normalized mass-to-flux ratio  of $1.8 \le \mu \le 23$.
The normalized mass-to-flux ratio is also listed in Table~\ref{table:1}.

\subsubsection{Sink Cell and Numerical Method}
\label{sec:sink}
To realize the long-term calculation of star formation, we adopt a sink at the centre of the cloud.
The detailed procedure for introducing sink cells is described in \citet{machida10a} and \citet{machida12}.
Here, we briefly describe the process.
We begin the calculation without a sink and calculate the cloud evolution for the pre-stellar gas collapse phase without a sink.
Then, we identify the protostar formation in the collapsing cloud when the number density exceeds $n > n_{\rm thr} =10^{13}\cm$ at the cloud centre.
After protostar formation, in the region $r < r_{\rm sink} = 2\,$AU, the gas with  a number density of $n > n_{\rm thr}$ is removed from the computational domain and added to the protostar as a gravitating mass in each time step.
Thus, for each time step, the accretion mass onto the protostar is calculated as
\begin{equation}
M_{\rm acc} = \int_{r < r_{\rm sink}} [\rho(i,j,k) - \rho_{\rm thr}]\, dV.
\end{equation}
We store the mass accretion rate in one-year increment.
It should be noted that because simulation timestep is  shorter than $\Delta t <0.01$\,yr, the mass accretion rate used in this study is averaged over $>100$ timesteps.

To calculate the outflow driving region and long-distance propagation of outflow, a wide range of spatial scale from $\sim1$\,AU (the scale of the circumstellar disk) to $\sim1$\,pc (the scale of the evolved protostellar outflow) must be covered.
To resolve such considerably different scales, we use the nested grid method  \citep{machida05a,machida05b}. 
Each level of a rectangular grid has the same number of cells ($ 64 \times 64 \times 32 $) and the grid size and cell width is halved for every increase in grid level.
The calculation begins with five grid levels ($l=1-5$).
The fifth level of the grid ($l=5$) has  a box size of $L_5=2\,r_{\rm BE} = 1.2\times10^4$\,AU.
Thus, the host cloud for models 1-8 is just embedded in the $l=5$ grid, while that for models 8 and 9 is embedded in the $l=4$ grid.
The first level of the grid has a box size of $L_1=2^5\,r_{\rm BE}=2.0\times10^5$\,AU (=0.97\,pc) filled with low-density interstellar medium $\rho_{\rm ISM}=0.01\rho_{0}$ outside $r>R_{\rm cl}$, where $\rho_{\rm ISM}$ is the density of the interstellar medium.
Thus, we are able to  calculate the propagation of the protostellar outflow in the region of  $< 2.0 \times10^5$\,AU.
However, the protostellar outflow never reached the computational boundary by the end of the calculation in any of the models.
After the calculation begins, a new finer grid is generated before the Jeans condition is violated \citep{truelove97}.
We set the maximum grid level of $l=12$ with a box size of 94\,AU and cell width of 1.46\,AU.
Thus, we are able to resolve the structure from $\sim1$\,AU to $\sim 1$\,pc.
We calculated the cloud evolution with various spatial resolutions, including various cell widths and grid sizes, to investigate the required spatial resolution for outflow driving. 
With these calculations, we checked the convergence of outflow momentum and energy and confirmed that the spatial resolution adopted in this study is sufficiently high to investigate the evolution of the outflow (\citealt{machida12}).

\subsection{Protostellar Calculation}
\label{sec:protostar}

We also calculate protostellar evolution
with variable mass accretion histories obtained through 
 the MHD simulations. 
We numerically solve the stellar structure equations by considering the effects of the mass accretion
\citep[e.g.][]{SST80,PS91}.
Our numerical codes have been developed in our previous
work to examine high-mass and low-mass protostellar
evolution with various accretion histories
\citep[e.g.][]{HO09,HYO10,HOK11}. 
In this paper, we calculate the protostellar
evolution separately from the MHD simulations, 
which give us accretion histories resulting from the interplay
between the powerful outflow and infalling envelope.


Our evolutionary calculation begins with a tiny initial model with a mass of $3 \times 10^{-3}~\msun$.
We follow the evolution in which the stellar mass increases
with a provided accretion history.
Mass accretion could significantly affect the stellar interior structure. 
An important and unknown quantity of the mass accretion
is the thermal efficiency, i.e. 
the specific entropy of gas settling onto the stellar surface.
In the stellar evolution calculations, the accretion thermal
efficiency is controlled by the outer boundary conditions of the models.
In this paper we adopt the shock outer boundary condition, 
which supposes that the spherical accretion flow directly
hits the stellar surface to form an accretion shock front 
\citep[e.g.][]{SST80}.
This condition corresponds to a relatively hot mass accretion because part of the entropy generated at the accretion shock
front is efficiently absorbed into the stellar interior.
Some authors argue that, in the low-mass star formation with an accretion rate of $\lesssim 10^{-5}~\mdot$, 
the accretion should be significantly colder.
However, the exact value of the accretion thermal efficiency remains highly uncertain, and this issue is beyond the scope of our study.
With the shock boundary, protostellar evolution is nearly independent of arbitrary initial models with unknown properties (such as mass and radius; e.g. \citealt{SST80b}).
With the outer boundary of the cold accretion, however, protostellar evolution 
differs with various  initial models 
as well as accretion histories \citep[e.g.][]{HCK97}.
In this paper, we adopt the shock boundary to focus only on
variations of protostellar evolution with various 
accretion histories. 
If the mass accretion was colder than the assumed value, the resulting
stellar radius would be relatively smaller than that determined by our results.


The protostellar calculations provides the evolution of the
total stellar luminosity
\begin{equation}
L_{\rm tot} = L_* + \frac{G M \dot{M}}{R} ,
\end{equation}
where the first and second terms on the right-hand side represent
the stellar and accretion luminosities, respectively. 
We refer to the total luminosity $L_{\rm tot}$ simply 
as protostellar luminosity hereafter.
In all of the examined cases, the accretion luminosity $L_{\rm acc}$
dominates the stellar luminosity $L_*$.
Because the protostellar evolution advances in the 
accretion timescale $M/\dot{M}$ in this case,
we update the stellar models each time the stellar
mass increases by $\sim 1$~\% with the accretion rate 
averaged over the timestep.
The very short accretion variability 
is eliminated through this procedure, however, the stellar structure change are minimal over such short periods.

\section{Results}
\label{sec:results}

\subsection{Classification of Evolutionary Stage}
\label{sec:class}

We calculated the cloud evolution for ten models with different
cloud parameters, as listed in Table~\ref{table:1}.
To characterize the evolution and to compare our results with observations, we defined three different evolutionary stages, Class 0, I, and II stages, using the envelope mass as
\[ \left\{
\begin{array}{l}
  {\rm Class \, 0\ stage: \ \ \ }   (M_{\rm env} / M_{\rm env,0}) \ge 0.5         , \\
  {\rm Class \, \, I\ stage: \ \ \ }   0.1  \le (M_{\rm env}/M_{\rm env,0}) < 0.5 , \\ 
  {\rm Class \, II\ stage: \ \ \ }  ( M_{\rm env}/M_{\rm env,0} ) < 0.1, 
\end{array}
\right.
\]
where $M_{\rm env,0}$ is the initial cloud mass, and the envelope mass  $M_{\rm env}$ is defined as 
\begin{equation}
M_{\rm env} = M_{\rm total}\, (r<\rcri) - M_{\rm disk},
\label{eq:envmass}
\end{equation}
where $M_{\rm total}\,(r<\rcri)$ is the total mass in the region of $r<\rcri$, and $M_{\rm disk}$ is the mass of the rotation-supported disk identified according to the prescription in \cite{machida10a} (see also \citealt{machida12}).
Prior to the formation of the protostar and rotating disk, the envelope mass $M_{\rm env}$ coincides with the initial cloud mass $M_{\rm env,0}$.
It should be noted that the protostellar mass $M_{\rm ps}$ is not included in equation~(\ref{eq:envmass}) because the gas composing the protostar is removed from the computational domain, as described in \S\ref{sec:sink}.
In addition, the mass of the host cloud with a radius of $r=\rcri$ (see \S\ref{sec:setting}) differs  from the envelope mass by the mass of the rotation-supported disk.
The gas falling onto the sink converts into a protostar and is removed from the computational domain, and part of the gas is expelled from the host cloud by the protostellar outflow.  
Therefore, the envelope mass never increases but decreases with time.

In this paper, we defined Class 0, I and II stages by using only the envelope mass.
With observational results, \citet{andre93} originally defined Class 0 protostars as objects which have $M_{\rm ps}/M_{\rm env}<1$.
This indicates that the envelope still retains approximately a half of the initial cloud mass during the Class 0 stage.
Thus, we defined the transition period between the Class 0 and I stages at the point at which 50\% of the initial cloud is accreted onto the protostellar system (protostar plus the circumstellar disk).
It should be noted that, in our definition, the envelope mass includes the outflowing gas in addition to the infalling gas.
We can determine this transition period using only the mass of the infalling gas.
However, we believe that our definition is plausible for comparing calculation results with observations because it is difficult to separate the infalling gas from the outflowing gas in observations.

The use of only envelope mass creates difficulties in determining the transition period between Class I and II stage because these classes are observationally determined through  spectral energy distribution.
However, such a method is useful in defining the classes with a single parameter ($M_{\rm env}$).
In this paper, we focus mainly on the evolution during the Class 0 and I stages, and briefly comment on the Class II stage.
In addition, it is considered that nearly all of the infalling envelope has already disappeared in the Class II stage.
Thus, we roughly define the transition period between the Class I and II stage as the period at which the envelope mass reaches 10\% of the initial cloud mass, as previously described (see also \citealt{vorobyov06}).

Table~\ref{table:2} summarizes the calculation results for each model.
The mass of the protostar ($M_{\rm ps,0}$), rotating disk ($M_{\rm disk,0}$) and protostellar outflow ($M_{\rm out,0}$) at the end of the Class 0 stage are listed from the second to fourth columns, respectively, of Table~\ref{table:2}.
The protostellar luminosity ($L_{\rm ps,0}$) at the end of the Class 0 stage is listed in the fifth column.
The duration ($t_0$) of the Class 0 stage after protostar formation is described in the sixth column.
The same quantities ($M_{\rm ps,I}, M_{\rm disk,I}, M_{\rm out,I}, L_{\rm ps,I}$ and $t_I$) at the end of the Class I stage are listed in the seventh to eleventh columns.
These quantities are not described for models 1, 4, 5, 9 and 10 because the calculation did not proceed in the Class II stage in such  models.
The mass of the protostar ($M_{\rm ps,II}$), rotating disk ($M_{\rm disk,II}$), protostellar outflow ($M_{\rm out,II}$) and envelope ($M_{\rm env}$) at the end of the calculation are listed in the twelfth to fifteenth columns.
In this section,  we describe the evolution of the cloud and protostellar outflow for a typical model (model 3) in \S\ref{sec:typical}, and we summarize the results for all models in \S\ref{sec:parameters}.

\subsection{Evolution of Protostellar Outflow for Typical Model}
\label{sec:typical}

\subsubsection{Structure and Collimation of Protostellar Outflow}
\label{sec:structure}

In this subsection, we show the cloud evolution for a typical model (model 3).
At the initial state, the host cloud has a radius of $r= 6.1 \times 10^3 $\,AU and a mass of $M=1.05\msun$.
The initial magnetic field strength and rotation rate are
$B_0 = 25\mu$\,G and $\Omega_0 = 1.0\times10^{-13}$\,s$^{-1}$, respectively.
The mass-to-flux ratio is $\mu=7$.
Observations show that  molecular cloud cores have a mass-to-flux
ratio of $0.8\lesssim \mu \lesssim7.2$ with a median value of $\lambda
\approx 2$ \citep[e.g.][]{crutcher99}.
Thus, Model 3  has a somewhat weaker magnetic field 
than that of the observed values.
The initial rotational energy of the cloud is $\beta=0.01$.
Observations show that molecular cloud cores have $10^{-4}<\beta_0 < 1.4$  with a typical value of $\beta_0 \sim0.02$ \citep[e.g.]{goodman93,caselli02}. 
The adopted rotational energy is also slightly smaller than that of the
observational estimates.

\begin{figure}
\includegraphics[width=150mm]{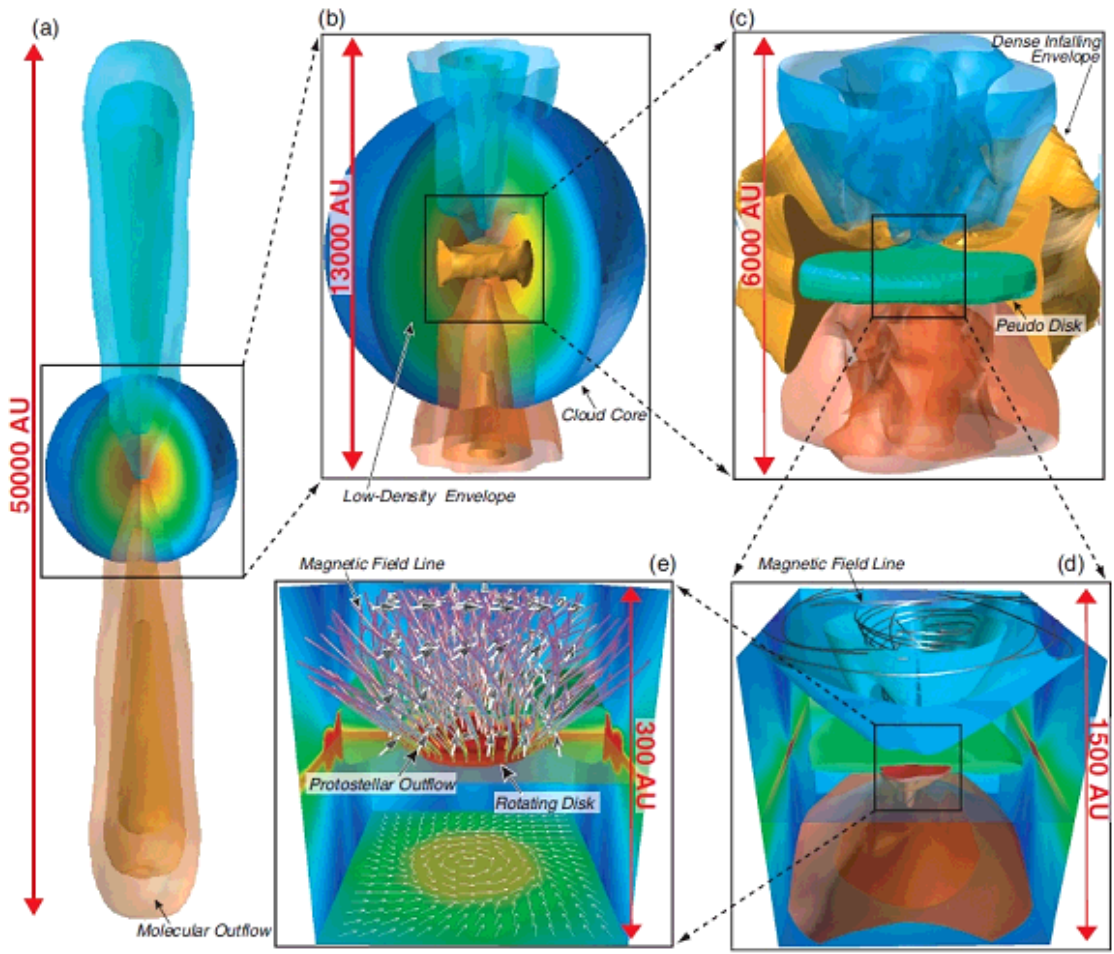}
\caption{
Structure of protostellar outflow at $t_{\rm ps}=1.447\times10^5$\,yr after the protostar formation with different scales for model 3.
The box scale is described in each panel.
The elongated blue and red structure in panels ({\it a})-({\it d}) mean the protostellar outflow inside which the gas is outflowing from the central region.
The central sphere with blue surface in panels ({\it a}) and ({\it b}) corresponds to the host cloud.
The colour on the cutting plane inside the host cloud means the density distribution.
The pseudo disk, infalling envelope and rotating disk are indicated by an arrow in each panel.
The thick arrows with black and white colour in panel ({\it e}) are velocity vectors in the outflowing region.
The magnetic field lines inside the $z>0$ outflowing region are plotted by streamlines in panels ({\it d}) and ({\it e}).
}
\label{fig:1}
\end{figure}

Figure~\ref{fig:1} depicts several key objects for guiding the
evolution of protostellar outflow in various spatial scales.
In this paper, we used three different elapsed times: the elapsed time after the cloud begins to
collapse ($t$), that after protostar formation ($t_{\rm ps}$) and
that after outflow emergence ($t_{\rm out}$). 
Figure~\ref{fig:1} shows the structure at $t=2.178\times10^5$\,yr
(Class II stage), which corresponds to $t_{\rm ps} = 1.447\times10^5$\,yr 
and $t_{\rm out}=1.473\times10^5$\,yr.
The protostar formation epoch $t_{\rm ps,0}$ and outflow emergence time
$t_{\rm out,0}$ are listed in Table~\ref{table:3}.


As indicated in Figure~\ref{fig:1}{\it a}, the protostellar outflow has a size of $\sim 2.5\times 10^4$\,AU on one side at this epoch.
Because the radius of the host cloud is $\rcri = 6.1\times 10^3$\,AU, the vertical length of the protostellar outflow is approximately four times the host cloud radius.
However, the horizontal width of the protostellar outflow is comparable to the host cloud radius (Fig.~\ref{fig:1}{\it a}).
\cite{machida12} reported that the width of the outflow reflects the host cloud radius because the protostellar outflow is anchored by the cloud-scale magnetic field lines and can widen up to the cloud scale.
Thus, we can observationally determine the size and mass of the core
from the (maximum) width of the protostellar outflow.
The blue sphere in Figures~\ref{fig:1}{\it a} and {\it b} corresponds to the host cloud.
In Figures~\ref{fig:1}{\it b} and {\it c}, inside the host cloud, the dense infalling envelope has a torus-like structure and can be identified by the orange iso-density surface of $n = 7\times10^4\cm$.
A pseudo-disk that is not rotation-supported disk is evident within the 
dense infalling envelope and is represented by the green 
iso-density surface.
As indicated  in Figures~\ref{fig:1}{\it c}, {\it d}, and {\it e}, 
the protostellar outflow is launched from a rotating disk that is surrounded by the pseudo-disk
in the very centre of the core.
At this epoch, the rotating disk has a size of $\sim100$\,AU, and the mass of the protostar and rotating disk are  $M_{\rm ps}=0.45\msun$ and $M_{\rm disk}=0.11\msun$, respectively.
As shown in Figure~\ref{fig:1}{\it e}, the rotating disk is vertically penetrated by the magnetic field lines that are significantly inclined against the disk rotation axis corresponding to the $z$-axis. 
Figure~\ref{fig:1}{\it d} shows that inside the
protostellar outflow, the magnetic field lines are strongly twisted due
to the disk rotation.
In summary, at this epoch, the outflow driving region is embedded in
$r \lesssim 100$\,AU, while the outflow extends up to $\sim2.5\times10^4$\,AU.

\begin{figure}
\includegraphics[width=150mm]{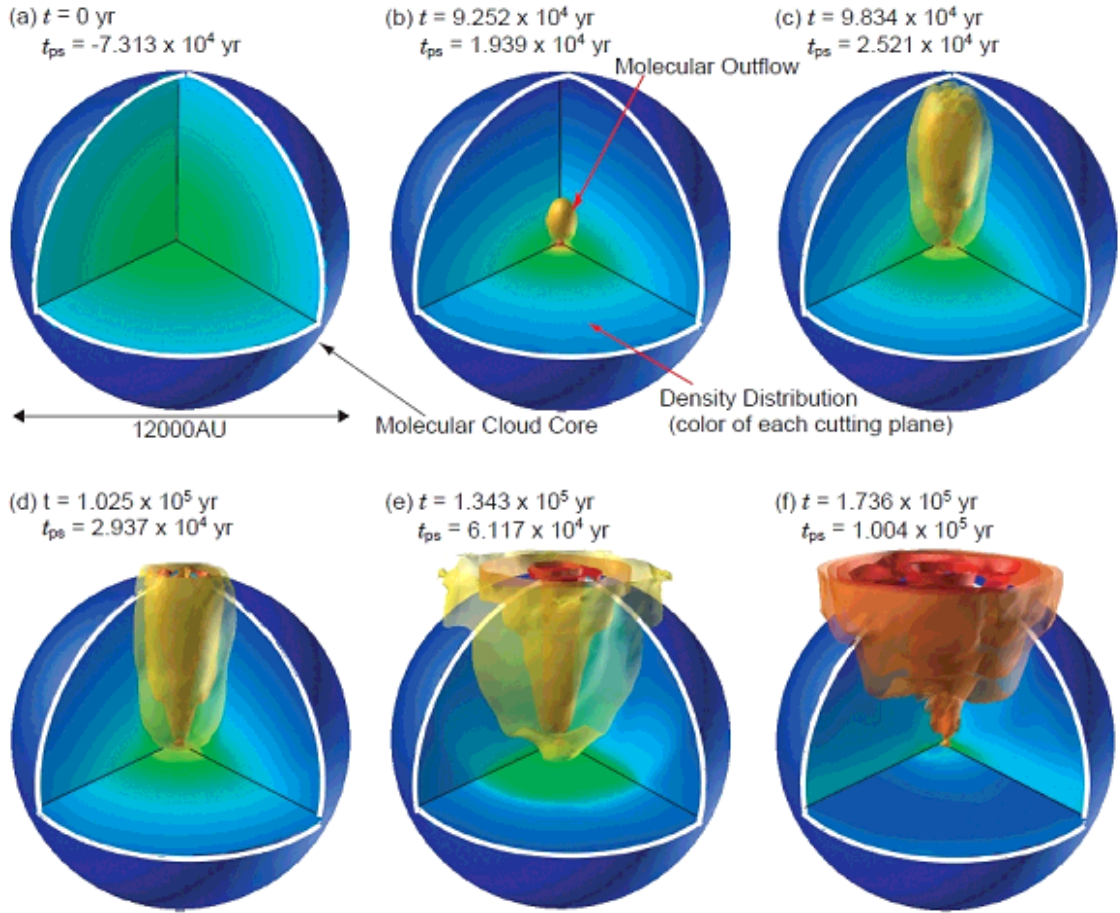}
\caption{
Time sequence of the protostellar outflow in a cloud scale for model 3.
The protostellar outflow is represented by yellow and orange iso-velocity surface inside which the gas has a positive radial velocity ($v_r>0$).
The inner host cloud region of $0^\circ < \theta < 90^\circ$ and $0^\circ < \phi < 90^\circ$ is plotted.
The colour on each wall is the density distribution on the cutting plane.
The elapsed $t$ and $t_{\rm ps}$ are described in each panel.
}
\label{fig:2}
\end{figure}

Figure~\ref{fig:2} shows the time evolution of the protostellar outflow for model 3 in the host cloud scale ($\sim 1.2\times10^4$\,AU).
To compare the protostellar outflow with the host cloud, we only show the region of $0^\circ < \theta < 90^\circ$ and $0^\circ < \phi < 90^\circ$ inside the host cloud, where $\theta$ and $\phi$ are the zenith and azimuthal angles, respectively.
Figure~\ref{fig:2}{\it a} shows the initial state for this model.
The protostellar outflow appears in the collapsing cloud
$7.047\times10^4$\,yr after the cloud begins to collapse. 
The outflow evolves and maintains a prolate shape,  as shown in Figures~\ref{fig:2}{\it b} and {\it c}.
The outflow reaches the boundary between the host cloud and interstellar medium at $t_{\rm ps}=2.521\times10^4$\,yr after protostar formation.
Then, the protostellar outflow penetrates the host cloud and flows into the interstellar space.
The gas is ejected from the host cloud by the protostellar outflow for $t_{\rm ps} > 2.521\times10^4$\,yr.

As shown in Figures~\ref{fig:2}{\it e} and {\it f}, the volume occupied by the protostellar outflow inside the host cloud gradually increases with time.
The protostellar outflow propagates along the magnetic field line.
The magnetic field lines open up as the distance from the equatorial plane increases and develops an hourglass-like configuration.
Thus, the opening angle of the protostellar outflow also gradually increases with increasing distance from the equatorial plane (for details, see \citealt{machida12}).
As a result, the protostellar outflow can sweep up and incorporate a large fraction of the infalling gas and eject it into the interstellar space.
Therefore,  protostellar outflow reduces the star formation efficiency \citep{nakano95,matzner00,machida12}.
After a large fraction of gas in the infalling envelope falls onto the protostellar system, or is expelled into the interstellar space (i.e. in the Class I or II stages), the volume occupied by the protostellar outflow becomes small.
The protostellar outflow shown in Figure~\ref{fig:2}{\it f} is slimmer than that in Figure~\ref{fig:2}{\it e} at its root, or near the outflow driving region because the protostellar outflow weakens as the infalling envelope is depleted.
Thus, during the Class I and II stages, the mass ejection rate from the host cloud gradually decreases.

\begin{figure}
\includegraphics[width=150mm]{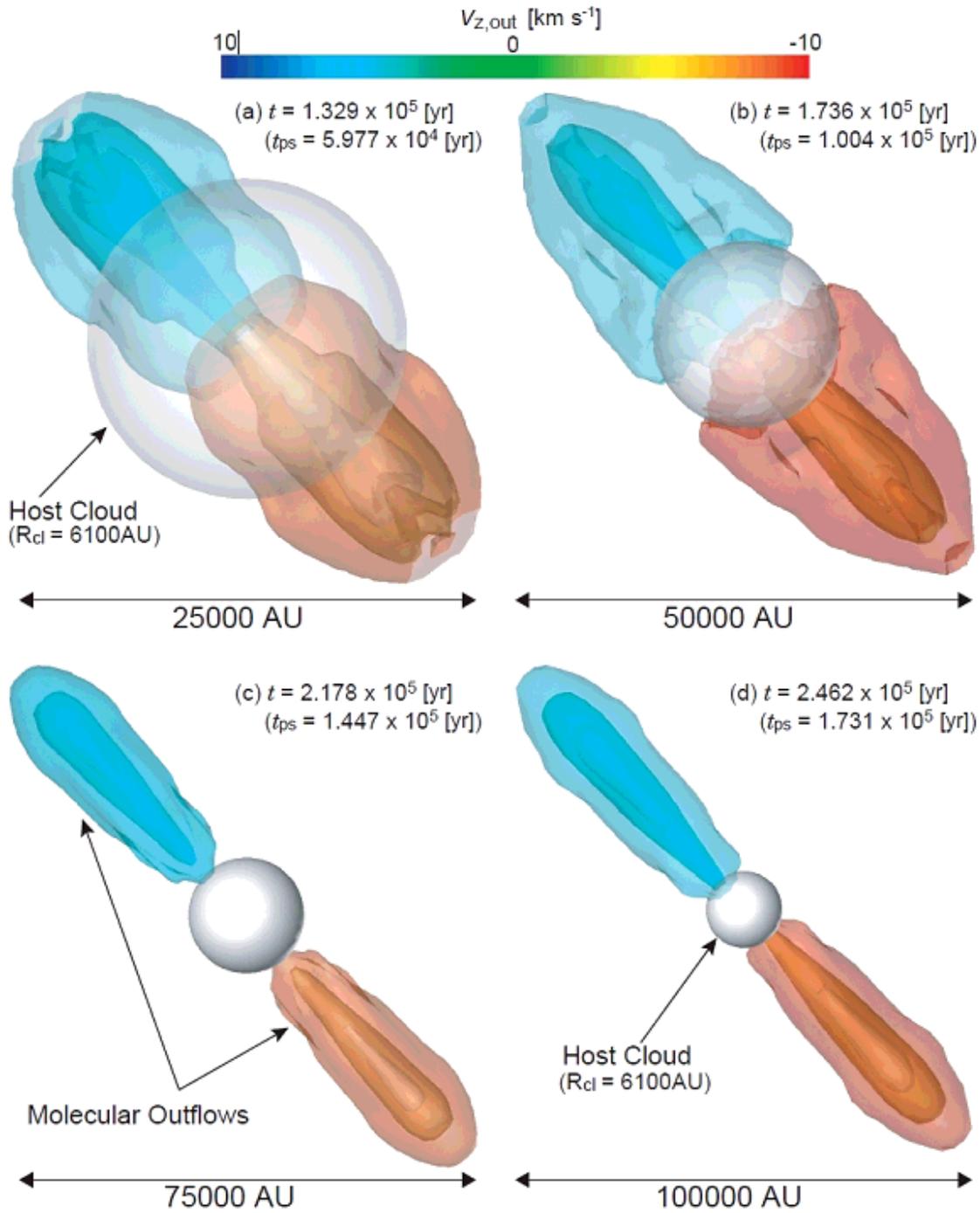}
\caption{
Evolution of the protostellar outflow in a large scale for model 3.
In each panel, the central grey sphere is the host cloud with a radius of $r=6.1\times10^3$\,AU.
The scale of each panel is different.
The outflow colour means the $z$-component of the outflow velocity, $v_{\rm z}$.
}
\label{fig:3}
\end{figure}

After the protostellar outflow breaks out of the cloud, it propagates into the interstellar space.
Figure~\ref{fig:3} shows the evolution of the protostellar outflow after the outflow vertical length exceeds the size of the host cloud;  the  scale of each panel is different to describe the entire region of the outflow.
The protostellar outflow is loosely collimated and has a relatively wide opening angle just after it penetrates the host cloud (Fig.~\ref{fig:3}{\it a}).
Then, the collimation of the outflow is gradually
improved with time (Figs.~\ref{fig:3}{\it b}-{\it d}) because the outflow extends only in the vertical direction and expands minimally in the horizontal direction.
As shown in Figure~\ref{fig:3}, the outflow always has a width comparable to the size of the host cloud.
As a result, the well-evolved outflow is well collimated
(Figs.~\ref{fig:3}{\it c} and {\it d}).
Such a well-collimated outflow is often observed in the star forming region \citep[e.g.][]{hirano06,velusamy07}.
As shown in Figure~\ref{fig:3}{\it d}, the protostellar outflow for model 3 reaches up to $\sim10^5$\,AU at the end of the calculation.

\begin{figure}
\includegraphics[width=150mm]{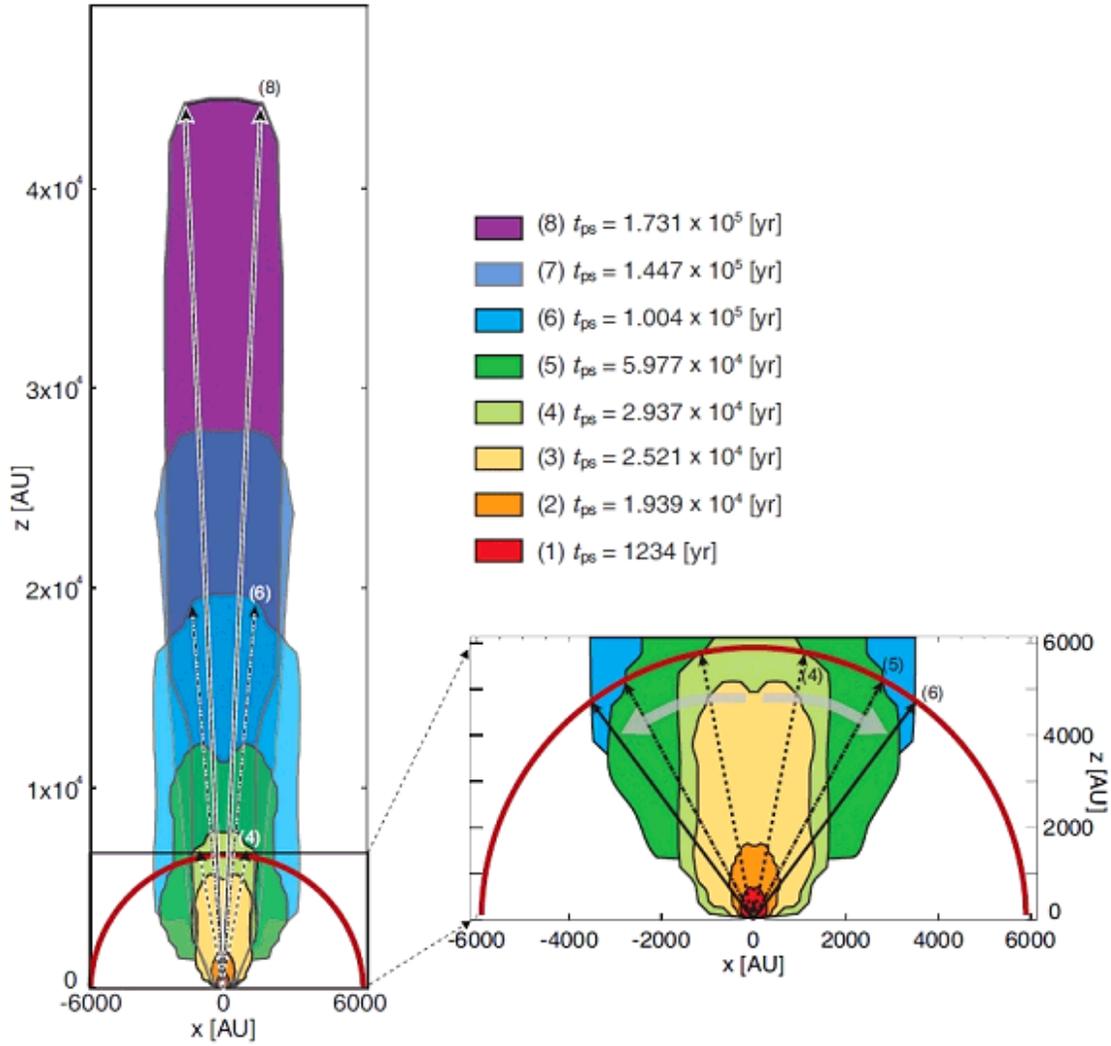}
\caption{
Shape of the outflow in the region of $z>0$ with different epochs for model 3.
Each epoch (the elapsed time after the protostar formation) is described in the upper right region.
The angle between two arrows in the left panel mean the outflow opening angle at each epoch, while that in the lower right panel is the opening angle in a cloud scale.
The grey arrows in the lower right panel is the evolutionary direction of the outflow opening angle in a cloud scale.
The red circle corresponds to the host cloud.
}
\label{fig:4}
\end{figure}

To investigate the morphological evolution of the protostellar outflow,
we present the shape of the outflowing region at each epoch 
in Figure~\ref{fig:4}.
As shown in the lower right-hand panel, before the
outflow reaches the cloud boundary 
($t_{\rm ps} \lesssim 3\times10^4$\,yr, epochs [1] - [4]), 
the outflow evolves and maintains nearly  the same ratio of
vertical to horizontal length.
Thus, during this period, the outflow opening angle rarely changes, and
the outflow collimation is poor.
After the outflow reaches the cloud boundary (epochs [5] - [8]), the
outflow extends only in the vertical direction as shown in
the left panel of the figure.
The opening angle of the protostellar outflow 
gradually becomes smaller with time.
Therefore, the outflow has a relatively wide opening angle in the earliest
evolutionary stage, while a well-collimated outflow is realized in
the later stage. 


On the contrary, in a cloud scale, or a scale smaller than the cloud radius, 
the opening angle increases with time, as indicated by grey arrows over epochs [4] - [6]
in the lower right panel.
Because the observations usually focus on the dense outflowing 
gas that is located near, or inside the cloud core, the opening angle of the outflow may be observed to apparently
increase with time.
\citet{velusamy98} observed  outflow-infall interaction in IRS1 in B5
with $^{12}$CO and $^{18}$CO emissions and showed that the opening
angle of the outflow should increase with time.
\citet{arce06} also reported  that the outflow cavity widens as the
envelope mass decreases with different observations.
These observations are consistent with our results in the cloud scale.
Because the outflow has a wide opening angle in the cloud scale, 
it effectively limits the gas accretion onto the protostellar system.

\subsubsection{Envelope Mass and Outflow Momentum}
\label{sec:typical-momentum}

\begin{figure}
\includegraphics[width=130mm]{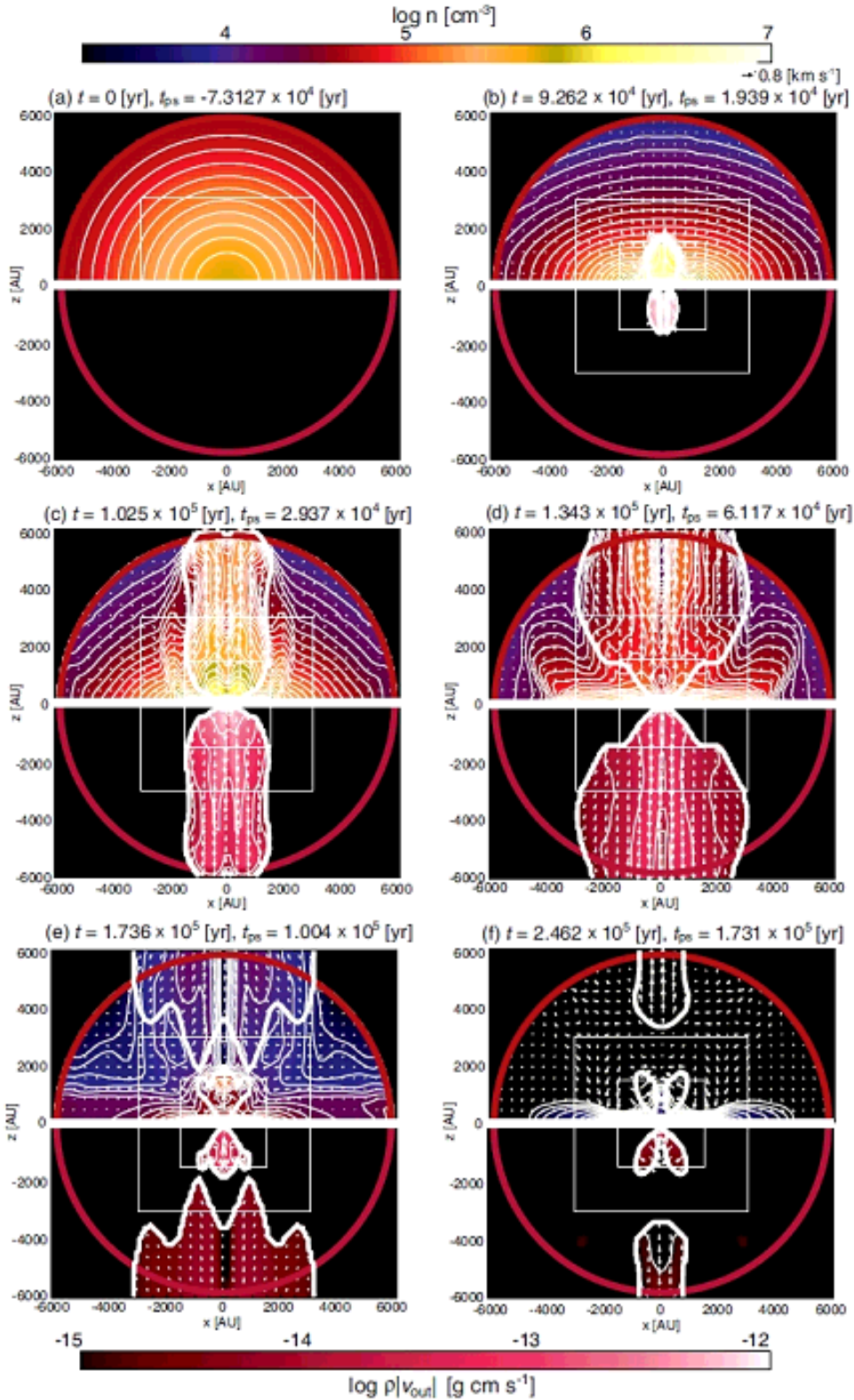}
\caption{
Evolution of the density (each upper panel) and outflow momentum (each lower panel) distribution on the $y=0$ cutting plane in the host cloud scale for model 3.
The white line in each panel corresponds to the boundary between the outflowing gas and infalling envelope.
The host cloud ($r=\rcri$) is plotted by the red circle.
Arrows are velocity vectors (only the velocity vectors in the outflowing region are plotted in each lower panel).
}
\label{fig:5}
\end{figure}

\begin{figure}
\includegraphics[width=130mm]{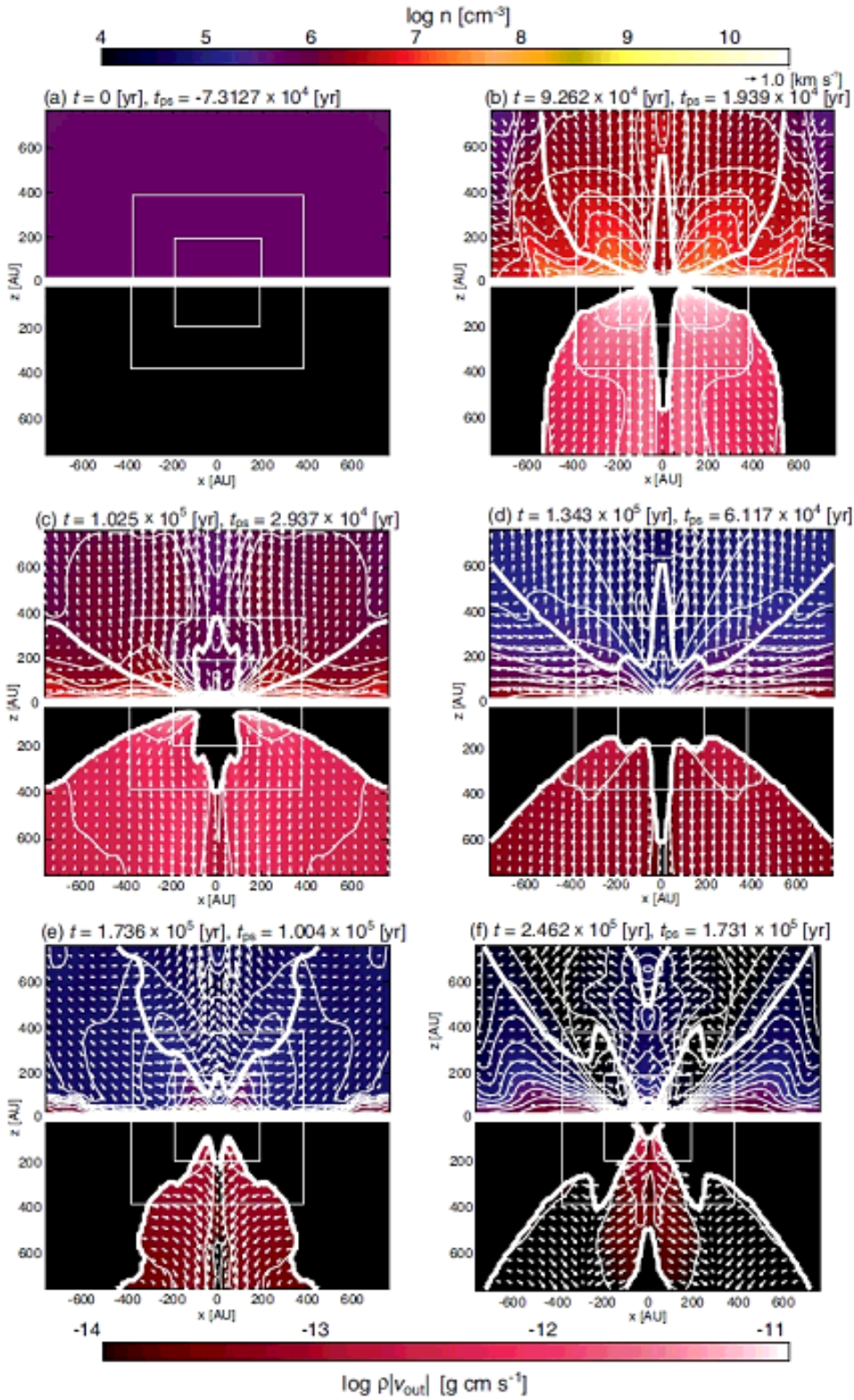}
\caption{
Same as in Fig.~\ref{fig:5}, but in the rotating disk scale.
}
\label{fig:6}
\end{figure}

The mass of the infalling envelope decreases with time as 
gas accretes onto the star and disk.
Because the outflow is powered by the mass accretion, 
the protostellar outflow ceases as the infalling envelope gets
depleted.
To illustrate this process, we show the density distribution of the infalling
envelope (each upper panel) and outflow momentum 
($\rho \vert v_{\rm out} \vert$; each lower panel) on the $y=0$ cutting
plane in Figures~\ref{fig:5} and \ref{fig:6}.
Figure~\ref{fig:5} shows the evolution of the entire host cloud with a box size of $1.2\times10^4$\,AU, while Figure~\ref{fig:6} shows the evolution of the infalling envelope around the outflow driving region with a box size of  $1500$\,AU.


As shown in the upper panels of Figure~\ref{fig:5}, the density of the infalling envelope decreases with time.
At the end of the calculation, the density just inside the host cloud $\rho( r = 6.1\times 10^4\,{\rm AU)}$ is much less than $10$\%  of that at the initial state.
As shown in Figures~\ref{fig:5}{\it c} and {\it d}, a dense infalling
envelope with $10^4\cm\lesssim n \lesssim 10^6\cm$ remains for
$t\lesssim 1.3\times10^5$\,yr, and the protostellar outflow has a
relatively large momentum.
As the gas density in the infalling envelope decreases, the outflow momentum also decreases, which is evident through a comparison of the panels in  Figure~\ref{fig:5} panels {\it c} and {\it d} with those in Figure~\ref{fig:5}{\it e} and {\it f}.
In addition, it is evident that the outflowing region is separated in
Figures~\ref{fig:5}{\it e} and {\it f}, which occurs because the protostellar outflow is intermittently driven by the
circumstellar disk in the later evolutionary stage 
($t\gtrsim 1.6\times10^5$\,yr).
It should be noted that  the outflow is continuously driven in the early stage ($t\lesssim 1.6\times10^5$\,yr; Figs.~\ref{fig:5}{\it b}-{\it d}).
At the end of the calculation, the envelope mass becomes less than $10$\% of the initial cloud mass, and the very weak outflow is driven by the circumstellar disk.


Figure~\ref{fig:6} shows that the outflow has a considerably wide opening angle near the equatorial plane during the main accretion phase.
The opening angle increases until $t\lesssim 1.5\times10^5$\,yr and has a maximum opening angle of $\sim130\degr$ at this scale.
Then, the opening angle shifts to decrease for $t\gtrsim1.5\times10^5$\,yr.
Because the outflow opening angle is large in the early evolutionary stage, the gas accretes onto the circumstellar disk only from the side as seen in Figure~\ref{fig:6}{\it b} and {\it c}.  
The mass accretion rate and outflow momentum decrease with the density of the infalling envelope (Fig.~\ref{fig:6}{\it d}). 
Then, the outflow is intermittently driven by the circumstellar disk and has a nested structure as shown in Figure~\ref{fig:6}{\it e} and {\it f}. 
The outflow finally disappears as most of the envelope gas 
is depleted \citep{machida12}.

\begin{figure}
\includegraphics[width=150mm]{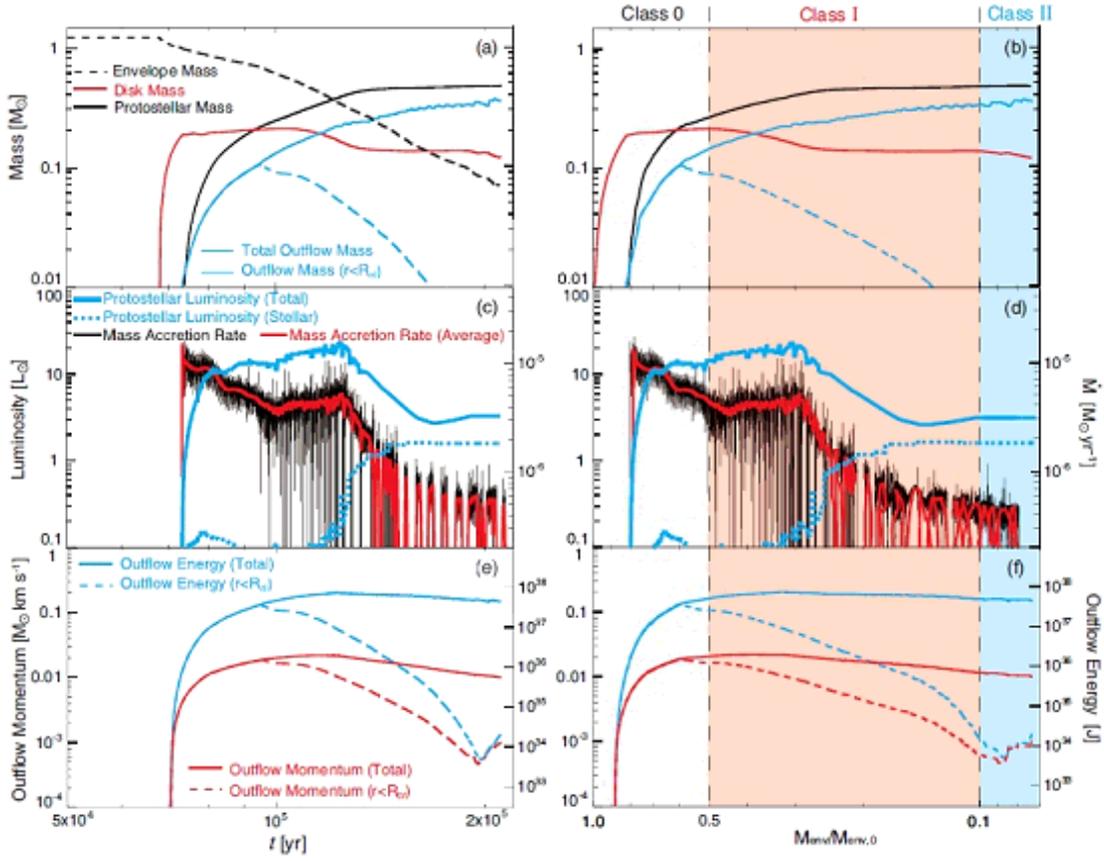}
\caption{
(a) Mass of protostar, disk, outflow and envelope, (c) total and protostellar luminosities (left axis) and mass accretion rate (right axis) and (e) outflow momentum (left axis) and energy (right axis) for model 3 are plotted against the elapsed time after the cloud begins to collapse.
The outflow mass in the host cloud ($r<\rcri$) and that in the whole computational domain are plotted in panel (a).
The momentum and energy of outflow in the host cloud ($r<\rcri$) and those in the whole computational domain are plotted in panel (e).
The same physical quantities are plotted against the envelope mass normalized by the initial host cloud mass in the right panels (b), (d) and (f).
}
\label{fig:7}
\end{figure}

The time evolution of the envelope mass is plotted against the elapsed time $t$ in Figure~\ref{fig:7}{\it a}, in which the mass of the protostar, rotating disk and outflow are also plotted.
The figure shows that the envelope mass is conserved for $t\lesssim
6.8\times10^4$\,yr.
At $t\sim 6.8\times10^4$\,yr,  the first (adiabatic) core, or the
rotating disk, forms in the collapsing cloud, and the mass of the infalling
envelope begins to decrease.
The first  core \citep{larson69,masunaga00} forms prior to the protostar formation and  is supported both by gas pressure and
rotation \citep{saigo06}.
The infalling gas continues to accrete onto the first core
until protostar formation.
After protostar formation, the first  core becomes the circumstellar
disk, which is  mainly supported by the rotation
\citep{bate98,bate11,machida10a,tsukamoto11}.
After protostar formation at $t_{\rm ps,0}=7.3127\times10^4$\,yr, 
the gas accretes onto the protostar
through the circumstellar, or rotating, disk.
Figure~\ref{fig:7}{\it a} shows that the circumstellar disk mass dominates the
protostellar mass for $t\lesssim 9\times10^4$\,yr, or approximately $2 \times 10^4$\,yr after the protostar formation.
The protostellar mass then dominates the circumstellar disk, which tends to become gravitationally stable.


The outflow is launched just prior to  protostar formation 
($t_{\rm out,0}=7.047\times10^4$\,yr).
As shown in \S\ref{sec:structure}, the outflow sweeps the gas in the
infalling envelope.
The protostellar outflow can eject a mass comparable to the protostellar mass into the interstellar space from the host cloud.
The blue dotted line in Figure~\ref{fig:7}{\it a} represents the outflowing mass inside the host cloud and indicates that the outflow gradually weakens for $t>10^5$\,yr.
The mass of total outflowing gas includes 
the outflowing gas within the host cloud, gas ejected from the host cloud and interstellar gas swept by the outflow.
The mass of the incorporated  interstellar gas is less
than $3$\% of the total outflowing mass because of the very low ambient density adopted
outside of the cloud (\S\ref{sec:setting}).


The masses of the protostar, disk and outflow are plotted against the
envelope mass normalized by the initial cloud mass in
Figure~\ref{fig:7}{\it b},
where the evolutionary stages of Class 0, I and II 
(see \S\ref{sec:class}) are denoted with background colours.
Figure~\ref{fig:7}{\it b} shows that the outflow mass inside the host cloud, represented by the blue broken line, shifts  to decreases just prior to the Class I stage.
In addition, the disk mass  gradually decreases during the Class I stage.
On the contrary, the protostellar and total outflowing mass slightly increase even during the Class I stage.

The mass accretion rate onto the protostar, represented by solid black line, is plotted
against the elapsed time $t$ in Figure~\ref{fig:7}{\it c} and against
the envelope mass ratio in Figure~\ref{fig:7}{\it d}, 
where the red line represents the mass accretion rate averaged over 1000\,yr.
Just after the protostar formation, the mass accretion rate is $\sim 10^{-5}\mdot$ and gradually decreases during the Class 0 stage.
Then, early in the Class I stage,  the averaged mass accretion rate temporally increases for $10^5\,{\rm yr}\lesssim t \lesssim 1.5\times10^5\,{\rm yr}$, or $0.3\lesssim M_{\rm env}/M_{\rm env.0} \lesssim 0.45$.
It is evident that the mass accretion onto the protostar is highly time variable.
This time variability is due to  the disk instability.
During this period, the circumstellar disk grows to become
gravitationally unstable and the angular momentum is
transferred by the gravitational torque in the disk in addition to the magnetic effects of protostellar outflow and
magnetic braking  \citep{machida10a,inutsuka10}.
The angular momentum transfer by the magnetic effects leads to steady
accretion, while that by disk instability often causes time
variable accretion \citep{vorobyov06,machida11a}.


After the infalling envelope mass decreases considerably, the gas
intermittently accretes onto the protostar from the circumstellar disk
($t\gtrsim1.5\times10^5$\,yr and $M_{\rm env}/M_{\rm env,0}\lesssim
0.3$).
Even during this phase, the gas continuously settles onto the
circumstellar disk from the infalling envelope with relatively low
mass accretion rates. 
The disk becomes gravitationally unstable with this continuous
mass supply from the infalling envelope.
The gravitationally unstable disk instantaneously amplifies the gas
accretion rate onto the protostar, which is known as the burst phase \citep{vorobyov06}.
Then, the disk mass decreases, and the mass accretion onto the protostar
temporally halts after the disk recovers to a stable state, known as the quiescent
phase \citep{vorobyov06}.
As shown in Figures~\ref{fig:7}{\it a}-{\it d}, however, 
the protostellar mass hardly increases during such burst-like
accretion events because of the short durations.


The evolution of the protostellar luminosity is also presented
in Figures~\ref{fig:7}{\it c} and {\it d}. 
Following protostar formation, the protostellar luminosity 
increases with the stellar mass
and has an initial peak of $L_{\rm ps}\sim 20 L_{\odot}$ 
at $M_{\rm env}/M_{\rm env,0} \simeq 0.3$ in the Class I stage.
The protostar luminosity then decreases as the accretion rate decreases.
This decrease in the protostellar luminosity during the Class I stage may solve the luminosity problem \citep{kenyon90,evans09,enoch09}.
\citet{offner11} also investigated the protostellar luminosity with different main accretion models  and pointed out that a gradual decrease of the accretion rate can solve the luminosity problem.
At the end of the Class I stage, the protostellar luminosity is 
$L \simeq 3~L_\odot$.
It should be noted that the evolution of the protostellar luminosity does not 
show spiky features, which are seen in the accretion
history for $M_{\rm env}/M_{\rm env,0} \lesssim 0.3$, because the short variabilities are smeared out 
for calculating the protostellar evolution.
However, the stellar structure would not be changed
with resolution of this episodic accretion because the 
increase of the stellar mass during that time is miniscule.
Figures~\ref{fig:7}{\it c} and {\it d} also indicate that the accretion luminosity dominates the stellar luminosity $L_*$ for $M_{\rm env}/M_{\rm env,0} \gtrsim 0.3$ while it is comparable to $L_*$ for $M_{\rm env}/M_{\rm env,0}\lesssim0.3$.


In Figures~\ref{fig:7}{\it e} and {\it f}, the evolution of the outflow momentum and energy are represented by the red and blue lines, respectively.
In the figure, these quantities are separately presented over the entire outflowing region and only the inside of the cloud by solid and broken lines, respectively.
During the early Class 0 stage, both the outflow momentum and energy
increase. 
However, just prior to the Class I stage the outflow momentum and energy for the inside of the cloud begin to decrease because the accreting matter, or the envelope mass, that
provides the driving force for the outflow is depleted during the Class I
stage.
At the end of the Class I stage, the outflow momentum and energy inside 
the cloud are approximately three orders of magnitude smaller than their peak values.
No powerful outflow is driven by the circumstellar disk during the Class
II stage. In this stage, the ejected outflow just propagates to nearly retain its original momentum and energy.
However, these quantities gradually decrease with time 
through interaction between the outflow and interstellar medium.

\subsection{Evolution of Protostellar Outflow in Clouds with Different Parameters}
\label{sec:parameters}
\subsubsection{Mass Evolution}
\label{sec:mass-evolution}
\begin{figure}
\includegraphics[width=150mm]{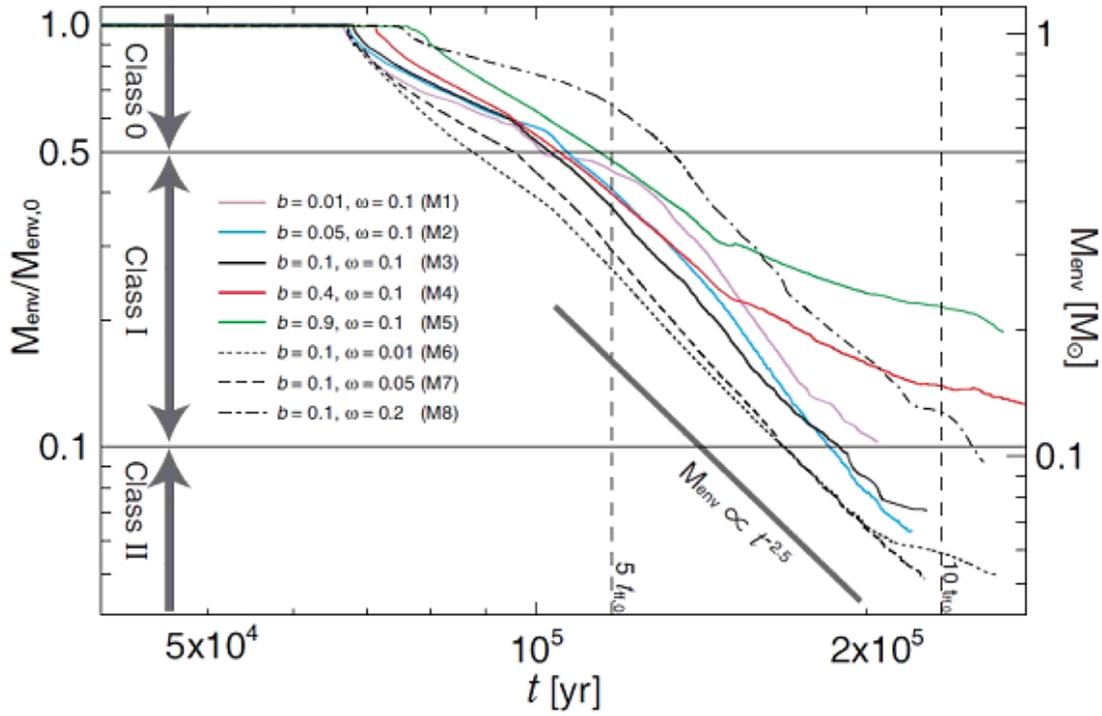}
\caption{
Time evolution of the envelope mass for models 1-8.
The left axis is the envelope mass normalized by the initial cloud mass, and right axis is the cloud mass in unit of $\msun$.
The evolutionary stages (Class 0, I and II) are described according to the classification in \S\ref{sec:class}. 
The elapsed time $t=5\,t_{\rm ff,0}$ and $10\,t_{\rm ff,0}$ are plotted by dash lines, where $t_{\rm ff,0}$ is the free fall timescale of the initial cloud.
The relation $M_{\rm env}\propto t^{-2.5}$ is also plotted by the thick line.
}
\label{fig:8}
\end{figure}

In this subsection, we explain cases with 
different cloud parameters.
Figure~\ref{fig:8} presents the evolution of the envelope mass, which is
normalized by the initial cloud mass, for models 1-8 against the time
after the cloud begins to collapse.
All models show similar evolution of the envelope mass.
Prior to protostar and circumstellar disk formation, the envelope mass
is nearly constant.
The rotating disk forms and the envelope mass begins to decrease 
$\sim(7-8)\times10^4$\,yr after the cloud begins to collapse.
The envelope mass halves, and the Class 0 stage ends approximately 
$10^4$\,yr after the formation of the protostar or rotating disk.


This result indicates that the Class 0 stage lasts for $\sim10^4$\,yr.
In fact, the durations of the Class 0 stage $t_0$ in the examined models
are in the range of 
$2.0\times10^4\, {\rm yr} < t_0 < 8.9\times10^4\, {\rm yr}$
(Table~\ref{table:2}). 
These values are comparable to those observationally estimated in \citet{andre93} and \citet{andre94}.
Recent observations indicate that the Class 0 lifetime is longer than $10^4$\,yr.
\citet{enoch09} observed many embedded protostars in Perseus, Serpens and Ophiuchus and estimated a Class 0 lifetime of $1.7\pm0.3\times 10^5$\,yr with a relative number of Class 0 and I sources (see also \citealt{evans09}).
More recently, \citet{maury11} shows a Class 0 lifetime of $\sim4-9\times10^4$\,yr with observations of Aquila rift complex. 
Thus, the Class 0 lifetime derived in this study agrees well with observations.


After protostar formation, the envelope mass decreases with $M_{\rm
env}\propto t^{-2.5}$ ($\dot{M}_{\rm env} \propto -2.5\, t^{-3.5}$) as shown in Figure~\ref{fig:8}.
It is difficult to analytically derive the power of $-2.5$, or $-3.5$, because the envelope mass includes
a non-negligible mass of the outflowing gas
ejected from the host cloud (Fig.~\ref{fig:7}). 
However, this rapid decrease in the envelope mass indicates that the infalling gas dissipates in a short duration after the protostar formation, and the main accretion phase does not last for a lengthy period.
That is, the duration of the Class I stage is not much longer than that of Class 0 stage \citep{enoch09,evans09}.
As described in Table~\ref{table:2}, the durations of the Class I stage $t_{\rm I}$ are in the range of $7.1\times10^4\, {\rm yr} < t_{\rm I} < 1.1\times10^5\, {\rm yr}$.

Next, we comment on the parameter dependence of the evolution of the envelope mass.
Figure~\ref{fig:8} shows that the envelope mass for models with slow initial rotation rates or weak magnetic fields (models 1,2, 6, and 7) decreases rapidly, while it for model with an initially rapid rotation or strong magnetic field (models 3, 4, 5, and 8) decreases slowly.
Both the cloud rotation and magnetic field slow the cloud collapse and its evolution because they can support the cloud against gravity \citep{scott80,machida05a}.
Thus, clouds with rapid rotations or strong magnetic fields have relatively long lifetimes of the infalling envelope.

However, the difference in cloud lifetime among models is not significant.
As shown in Figure~\ref{fig:8}, each cloud dissipates in approximately $\lesssim 10\,t_{\rm ff,0}$ after cloud begins to collapse, where $t_{\rm ff,0}$ is the freefall timescale of the initial cloud.
Thus, independent of magnetic field strength and rotation rate, clouds with the same mass (or same central density)  have nearly the same lifetimes because  gravity and pressure gradient force mainly control the cloud evolution, while rotation and magnetic field offer minor contributions \citep{machida05a}.

\begin{figure}
\includegraphics[width=150mm]{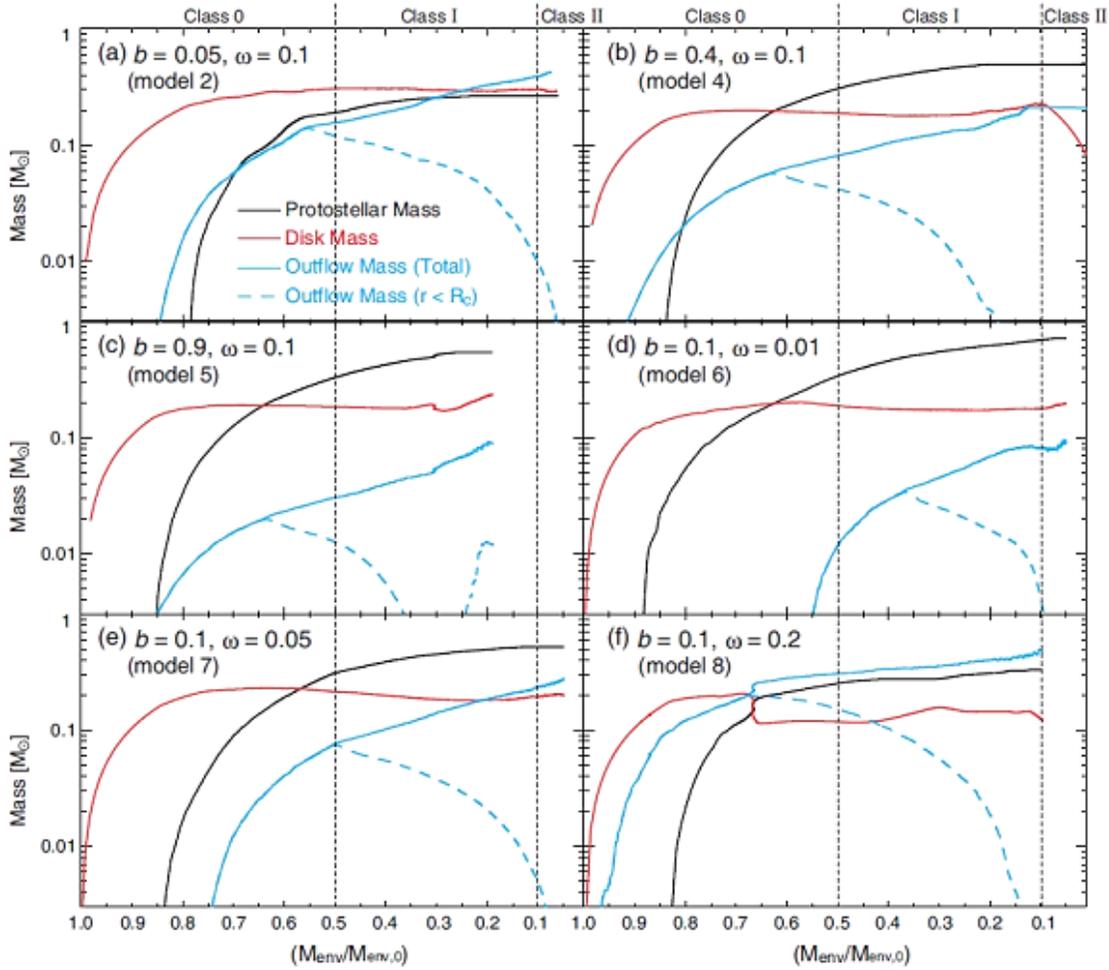}
\caption{
Mass of protostar, disk and outflow for models 2, 4, 5, 6, 7 and 8 are plotted against the envelope mass normalized by the initial cloud mass, respectively.
In each panel, the blue solid lines are the outflow mass in the whole computational domain, while the blue broken lines are the outflow mass inside the host cloud $r<\rcri$.
}
\label{fig:9}
\end{figure}

The envelope gas gradually accretes onto the circumstellar disk.
Then, part of the circumstellar disk gas is blown away by the protostellar outflow, and part of it accretes further onto the protostar.
The mass of the protostar, circumstellar disk and protostellar outflow for models 2, 4, 5, 6, 7 and 8 are plotted against the normalized envelope mass in Figure~\ref{fig:9}.
In the figure, although models have the same initial cloud mass (Table~\ref{table:1}), the mass ratio of each object, which includes the protostar, disk and outflow, differs considerably.
Furthermore, Figure~\ref{fig:9} indicates that  star formation efficiency ($\varepsilon$) strongly depends on cloud parameters $b$ (magnetic field strength) and $\omega$ (rotation rate), where the star formation efficiency is defined as the ratio of the protostellar mass to the host cloud mass $\varepsilon = M_{\rm ps}/M_{\rm env,0}$.
This result is expected because the outflow efficiency, which determines star formation efficiency, depends on the magnetic field strength and rotation rate of the cloud core.
When the host cloud has no magnetic field, all the envelope, or host cloud, mass accretes onto the protostellar system without emergence of the protostellar outflow. 
When the host cloud has no rotation and no circumstellar disk forms, then all of the envelope mass falls directly onto the protostar.
Thus, in the case with no rotation, we expect a star formation efficiency of $\varepsilon = 1$.
With rotation and magnetic field, the protostellar outflow can reduce the star formation efficiency
down to $\epsilon \sim 0.5$. 

As shown in Figure~\ref{fig:9}, for all models, the rotating disk appears prior to protostar formation \citep{bate98,bate11, walch09a,machida10a, tsukamoto11}. 
Just after protostar formation, the rotation disk mass dominates the protostellar mass \citep{inutsuka10}.
In the figure, the disk mass dominates the protostellar mass until the end of the calculation only for model 2 (panel {\it a}), while the protostellar mass dominates the disk mass as the envelope mass decreases in other models (models 4, 5, 6, 7 and 8).
The initial magnetic field for model 2 is relatively weak ($b=0.05$), and the angular momentum is not effectively transferred by the magnetic braking and outflow.
Thus, for model 2, a massive disk comparable to the protostellar mass remains even in the Class II stage.
Figure~\ref{fig:9} also indicates that a sufficiently large disk already exists in the Class 0 stage.
Recent observation confirmed a large circumstellar disk in the Class 0 stage \citep{tobin12}, which is consistent with our results.

Next, we focus on the mass of the protostellar outflow.
In models 2 (panel {\it a}) and 8 (panel {\it f}), the mass of the protostellar outflow is comparable to or  larger than the protostellar mass during the Class 0, I and II stage.
For these models, the initial cloud has a somewhat weak magnetic field and relatively rapid rotation.
When the host cloud is strongly magnetized, no powerful outflow appears, as shown by model 5 (Fig.~\ref{fig:9}{\it c}),  because the disk angular momentum is effectively transferred by the magnetic braking, and no massive disk to drive the protostellar outflow appears.
In addition, when the host cloud has a small angular momentum, neither a sufficiently large disk nor  powerful outflow appear (model 6, Fig.~\ref{fig:9}{\it d}).
Thus, the host cloud that has a moderately strong magnetic field and rapid rotation can drive a powerful, or massive, outflow, as seen in models 2 and 8.
However, even in models 5 and 6, the mass fraction of protostellar outflow is not negligible; at least $>10$\% of the host cloud mass is ejected from the host cloud by the protostellar outflow.

Figure~\ref{fig:9} indicates that, for all models except for model 6,  the outflow mass inside the host cloud has a peak during the Class 0 stage and decreases during the Class I stage.
In addition, as shown in the figure, the outflow appears prior to  protostar formation and after disk formation for models 2, 4, 5 and 8, while it appears after the protostar formation for models 6 and 7.
In general, the outflow is driven prior to protostar formation  \citep[e.g.][]{tomisaka02,banerjee06,machida08b,hennebelle08a,tomida10a, tomida10b,tomida12,duffin09,duffin11,price12,seifried12}.
The delayed emergence of the outflow for models 6 and 7 is due to the initial small rotation rate.
It is expected that, in these models, the outflow appears after the disk acquires a sufficient angular momentum from the infalling envelope.

Figure~\ref{fig:9} shows that the total outflow mass  is in the range of $M_{\rm out} \sim 0.1-0.5 \msun$.
Thus, approximately $10-50$\% of the host cloud mass is ejected by the protostellar outflow because the initial cloud has a mass of $M_{\rm cl}=1.05\msun$.
\citet{curtis10} investigated the outflow mass in four active star forming regions and showed that the observed outflows had masses in the range of $0.002\msun < M_{\rm out} < 0.4\msun$ with an average value of $0.09\pm0.02\msun$ for Class 0 protostars and $0.06\pm0.03\msun$ for Class I protostars.
Although we expect that the outflow mass depends on the initial cloud mass
(\S\ref{sec:host-cloud-mass}), our results roughly agree with the
observations.

\subsubsection{Outflow Momentum}
\label{sec:outflow-momentum}
\begin{figure}
\includegraphics[width=150mm]{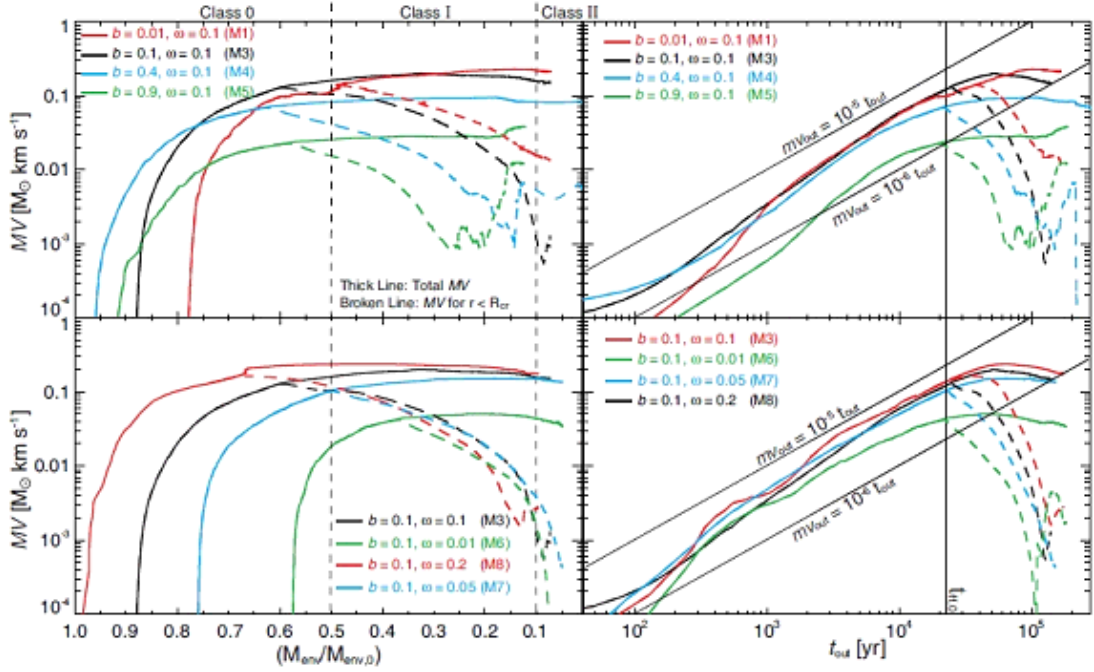}
\caption{
Outflow momentum for models with different initial magnetic field strengths (upper panels) and 
different initial cloud rotation rates (lower panels) against the envelope mass normalized by the initial cloud mass (left panels) and the time after the protostellar outflow emerges (right panels).
In each model, the outflow momentum in the whole region (solid line) and that inside the host cloud (broken line) are plotted.
}
\label{fig:10}
\end{figure}

The outflow momentum is a useful index to
observationally identify the evolutionary stage of a protostar
\citep{cabrit92, bontemps96, arce06,wu04,hatchell07,curtis10,andre00}.
In Figure~\ref{fig:10}, the outflow momenta for models 1, 3, 4, 5, 6, 7
and 8 are plotted against the normalized envelope mass (left panels)
and the elapsed time after the outflow appears (right panels).
The variations with different magnetic fields and
with different rotation rates are separately presented in the
upper and lower panels.
The outflows momenta are in the range of $0.02\,\mv < MV_{\rm out} <
0.2\,\mv$ at their peak.
\citet{curtis10} observed 45 outflows in the Perseus molecular cloud and
showed that the outflow momentum for Class I objects without a highly
powerful anomaly (SVS13) is $0.10\pm0.03 \mv$ on average.
Thus, the outflow momentum derived in our study is comparable to that of the observational estimates.
As shown in the left panels in Figure~\ref{fig:10}, the outflow momentum is 
larger with weaker magnetic fields
(models 1 and 3; upper panel) or with higher angular momentum
(models 3 and 8; lower panels). Both a moderately strong magnetic field and rapid rotation are necessary for driving a powerful outflow (see \S\ref{sec:mass-evolution}).


With the same initial rotation, 
the outflow momentum inside the host cloud begins to decrease
at nearly the same point $(M_{\rm env}/M_{\rm env,0}) \sim 0.6$
even with  different magnetic fields
(broken lines in Fig.~\ref{fig:10} upper left panel).
On the contrary, the duration for driving the powerful outflow
depends on the cloud rotation rate (broken lines in
Fig.~\ref{fig:10} lower left panel).
These results indicate that the total amount of the cloud angular momentum is strongly related to the duration of the powerful outflow driving.
This duration controls the total amount of the ejected mass 
and outflow momentum (solid lines in Fig.~\ref{fig:10}).


The right-hand panels in Figure~\ref{fig:10} show that outflow momentum is
linearly proportional to the elapsed time $t_{\rm out}$ after emergence of the
outflow. 
This evolution is approximately written as 
\begin{equation}
{MV_{\rm out}} 
= 5 \times 10^{-6} \left( \frac{t_{\rm out}}{{\rm yr}} \right)\, \mv.
\label{eq:momentum}
\end{equation}
Inside the host cloud ($r<\rcri$),  relation (\ref{eq:momentum}) can
be applicable for $t_{\rm out}<t_{\rm ff,0}$, which indicates that the duration of outflow driving 
is approximately equivalent to the freefall timescale of the host cloud. 
During this period, the gas is vigorously ejected around the
circumstellar disk.
After $t_{\rm out}>t_{\rm ff,0}$, the outflow momentum inside the host
cloud drastically decreases, and the outflow intermittently appears as
described in \S\ref{sec:typical-momentum}.

\subsubsection{Protostellar Luminosity and Mass Accretion Rate}
\label{sec:mass-accretion-rate}

\begin{figure}
\includegraphics[width=150mm]{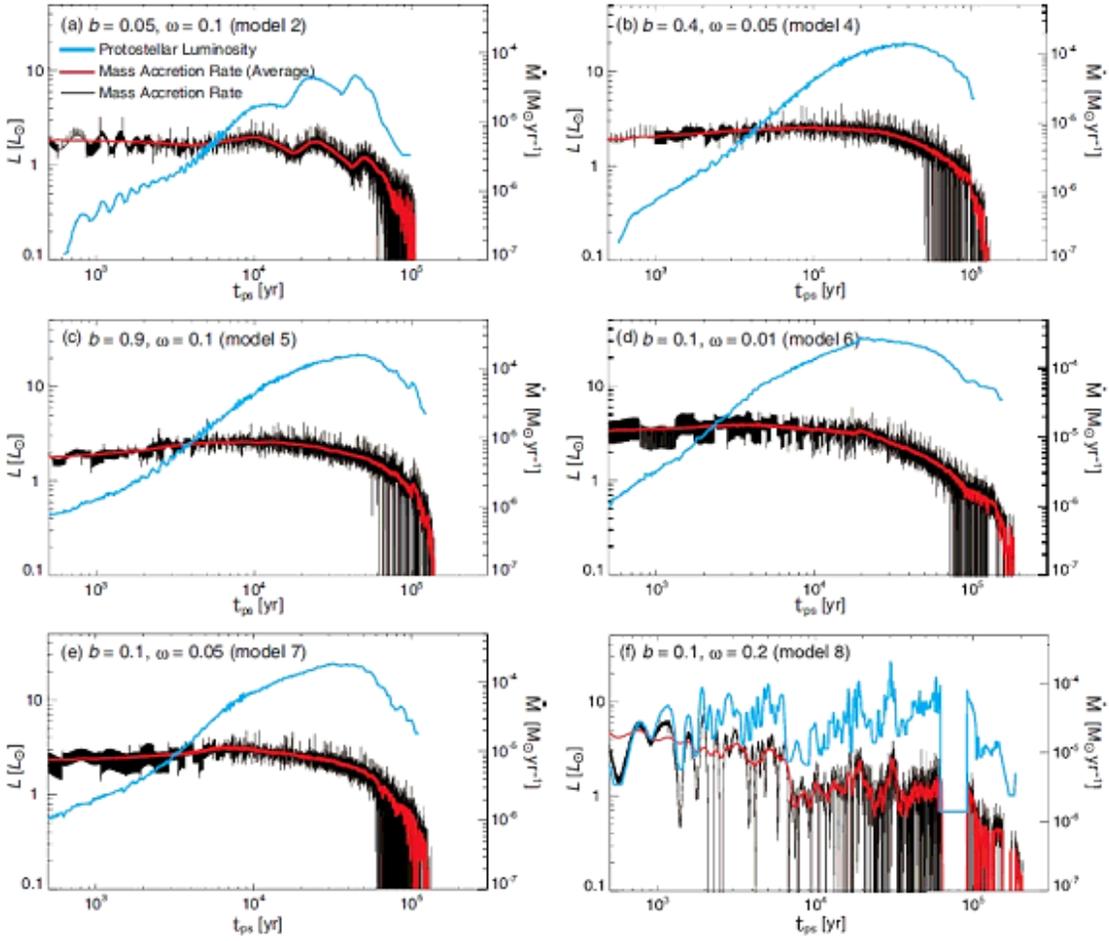}
\caption{
Protostellar luminosity (blue line; left axis) and mass accretion rate (black and red lines; right axis) for 
models 2, 4, 5, 6, 7 and 8 are plotted against the elapsed time after the protostar formation.
The red line in each panel is the mass accretion rate averaged every $1000$\,yr.
}
\label{fig:11}
\end{figure}

In addition to the outflow momentum, the protostellar luminosity 
could represent different stages of protostellar evolution.
Figure~\ref{fig:11} shows a time evolution of protostellar
luminosity and mass accretion rates onto the protostar
in models 2, 4, 5, 6, 7 and 8.
The figure shows  common accretion histories among the models
except for model 8; the accretion rate is nearly constant at 
$\dot{M}_{\rm ps} \sim10^{-5}\mdot$ for $t_{\rm ps} \lesssim10^4$\,yr
and gradually decreases after that time.
The mass accretion becomes time-variable for 
$t_{\rm ps} \gtrsim {\rm several} \times 10^4$\,yr and
almost ceases with $\dot{M}_{\rm ps} < 10^{-7}\mdot$ for 
$t_{\rm ps} \gtrsim 10^5$\,yr.
Reflecting such mass accretion histories, the protostellar luminosity gradually increases from $\sim 0.1\lsun$ to $\sim10\lsun$ for $t_{\rm ps} \lesssim5\times10^4$\,yr.
The luminosity then decreases as the mass accretion rate decreases for $t_{\rm ps}\gtrsim 5\times 10^4$\,yr.

\begin{figure}
\includegraphics[width=150mm]{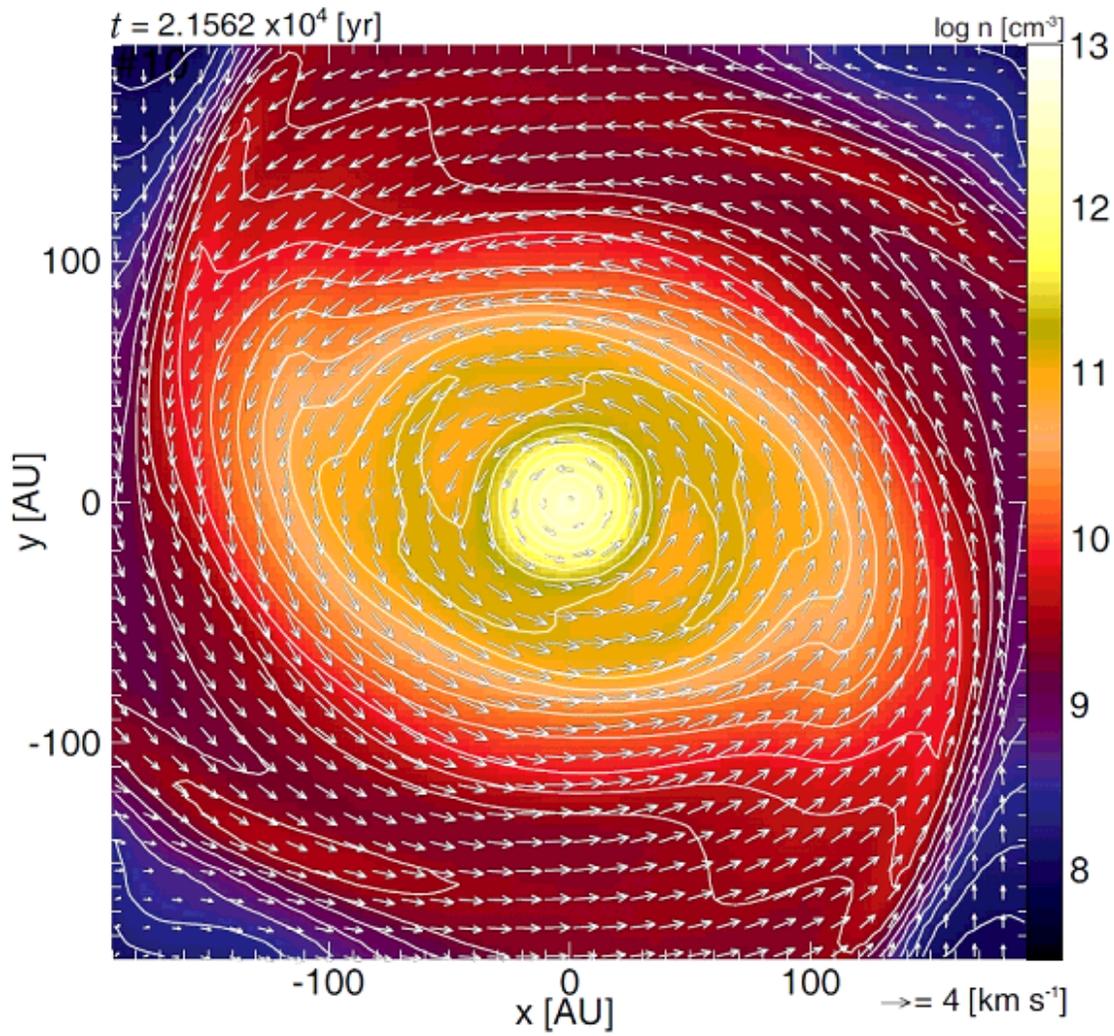}
\caption{
Density (colours and contours) and velocity (arrows) distributions around the protostar on the equatorial plane for model 8.
The elapsed time is described in the upper left corner.
}
\label{fig:12}
\end{figure}

Unlike models 2, 4, 5, 6 and 7, model 8 shows a highly time-variable
mass accretion throughout the evolution.
With the initial rapid rotation in this model, 
the circumstellar disk becomes highly gravitationally unstable, which
makes the accretion history significantly time-variable. 
Figure~\ref{fig:12} shows the density distribution in the disk for model 8.
The disk shows a non-axisymmetric structure
created by the gravitational instability, with which the
angular momentum is transported.
With the highly time-variable accretion, 
the protostellar luminosity is also significantly time-variable
throughout the accretion phase (Fig.~\ref{fig:11}{\it f}).
If the mass accretion appears through the highly gravitationally unstable
accretion disk, it would be thus difficult to characterize the
protostellar evolutionary stages simply from luminosity.
It should be noted that the odd mode of the non-axisymmetric density perturbation is suppressed in the disk because we fixed the sink (or protostar) at the center of the cloud as described in \S\ref{sec:sink}.
\citet{kratter10} showed that non-axisymmetric m=1 mode tends to develop and fragmentation occurs in the disk \citep[see also][]{tsukamoto11}.
In such a case, the mass accretion rate and protostellar luminosity may be somewhat different from those in Figure~\ref{fig:11}{\it f}. 

\begin{figure}
\includegraphics[width=100mm]{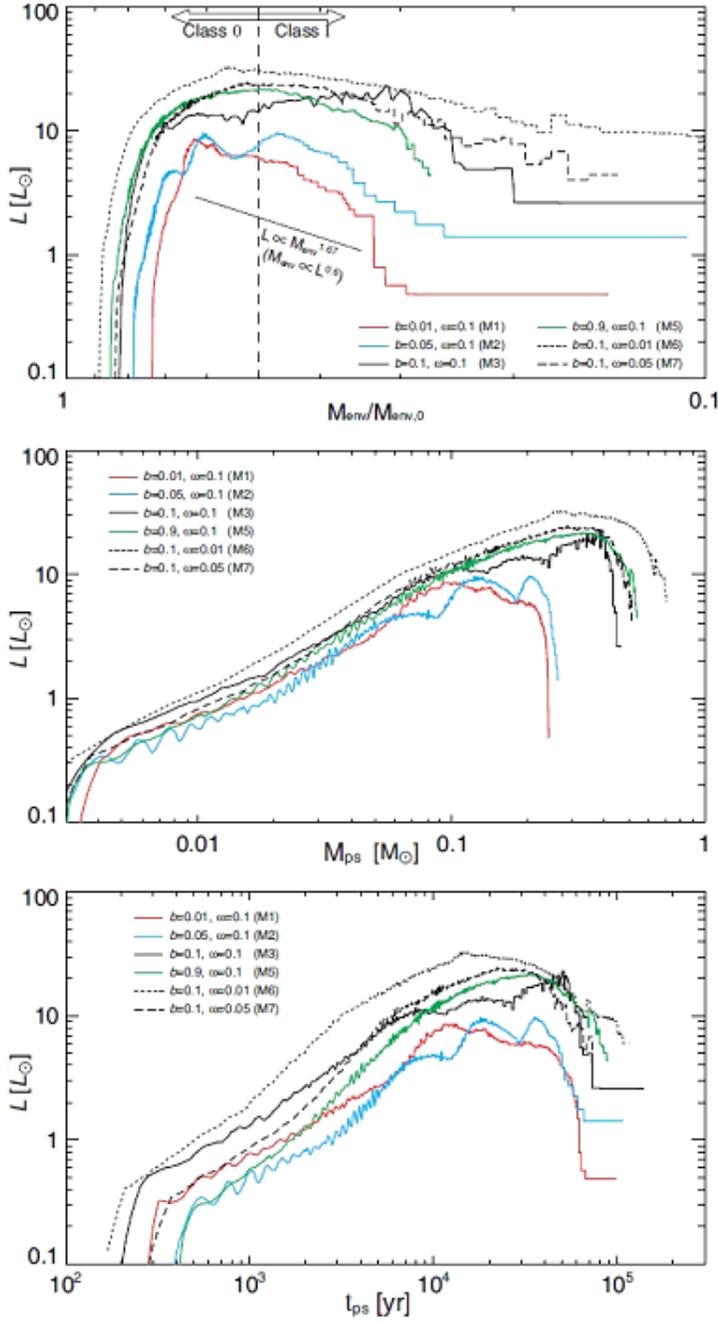}
\caption{
Protostellar luminosities for models 1, 2, 3, 5, 6 and 7 are plotted against the normalized envelope mass (upper panel), protostellar mass (middle panel) and elapsed time after the protostar formation (lower panel), respectively.
The relation of  $L \propto M_{\rm env}^{1.67}$ ($M_{\rm env} \propto L^{0.6}$) is plotted in upper panel.
}
\label{fig:13}
\end{figure}

Figure~\ref{fig:13} summarizes the evolution of protostellar
luminosity in the examined models against the normalized envelope 
mass (upper panel), the protostellar mass (middle panel) and 
the elapsed time after protostar formation (lower panel).
Model 8 is omitted here because its complex luminosity evolution (Fig.~\ref{fig:11}{\it f}) 
obstructs easy viewing of the evolutionary tracks.
In each model, the protostellar luminosity generally peaks in the Class 0 stage and decreases in the Class I stage.
At their peaks, the protostellar luminosities reach $\sim 8-30\lsun$.
Figure~\ref{fig:13} also shows that stellar luminosity is higher 
with  slower initial rotation (e.g. models 6 and 7) and lower 
with  weaker magnetic field (e.g. models 1 and 2), which reflects the fact that the mass accretion rate is relatively
lower with a more rapid rotation or weaker magnetic field.
The disk rotation suppresses the rapid gas accretion onto the protostar 
because the gas is supported by centrifugal force in the disk.
On the contrary, the magnetic field promotes gas accretion with an angular momentum transfer by magnetic braking.
Thus,  protostars forming with  slower rotation or stronger
magnetic fields have the higher protostellar luminosity.


The upper and lower panels in Figure~\ref{fig:13} indicate that in each model,
the protostellar luminosity during the (late) Class 0 stage does not differ significantly from that during the (early) Class I stage.
The observations also show a smooth transition of stellar 
luminosity from the Class 0 to I stages 
\citep[e.g. Fig.~1 of][]{bontemps96}.
On the contrary, at the same epoch,  the luminosity difference is as
large as one order of magnitude among models with different cloud
parameters (Fig.~\ref{fig:13} upper and lower panels).
This result indicates that, in observations, dispersion 
of the protostellar luminosity may be due to different properties 
of clouds such as magnetic field strength and rotation rate 
rather than to different evolutionary stages.

\section{Outflow Momentum Flux: Comparison with Observations}
\label{sec:observations}
\subsection{Momentum Flux vs. Bolometric Luminosity}

\begin{figure}
\includegraphics[width=150mm]{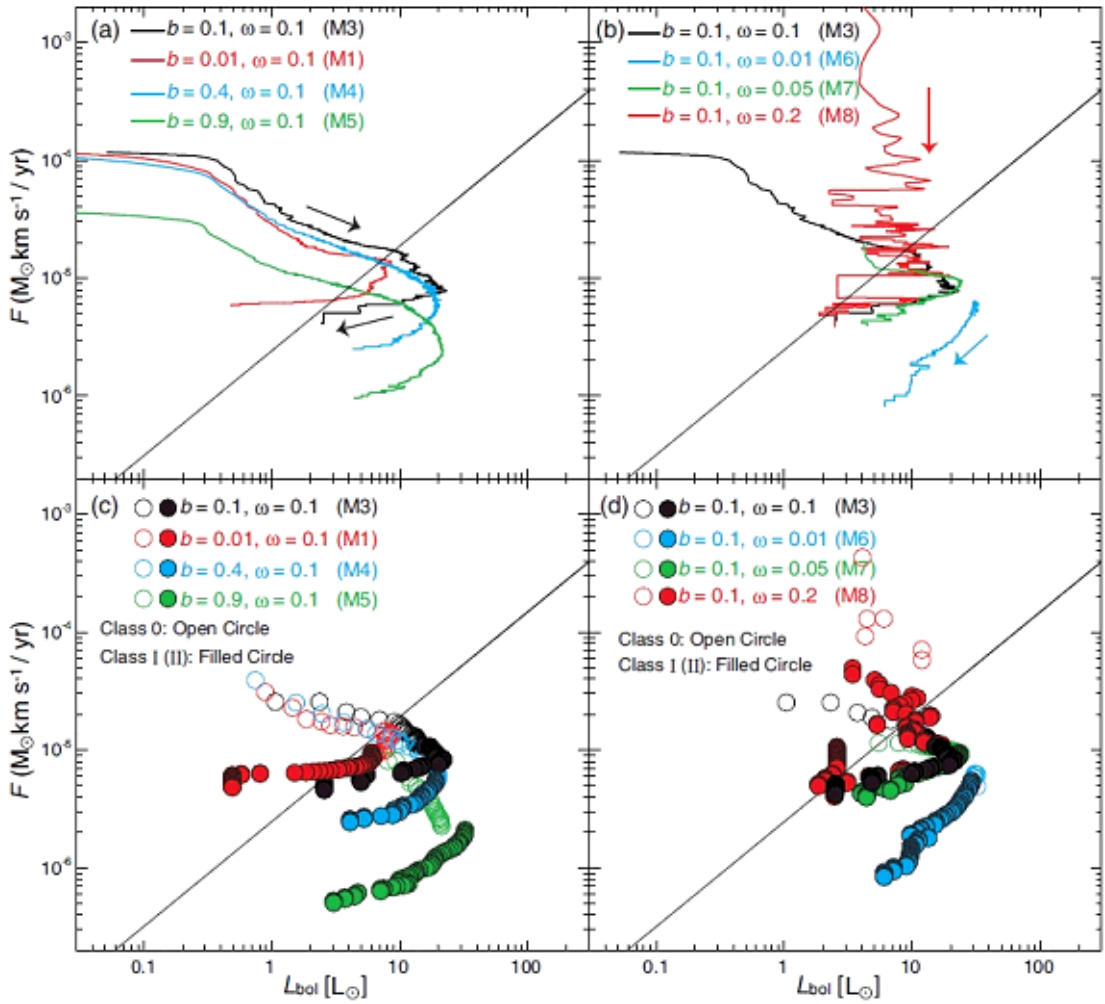}
\caption{
Outflow momentum fluxes ($F$) for models with different magnetic field strengths (left panels) and 
different initial cloud rotation rates against the protostellar luminosity are plotted in the upper panel.
The outflow momentum flux every 1000\,yr are plotted in the lower panels.
The open and filled circle mean the momentum flux during the Class 0 and I stages, respectively.
The plotted ranges and solid line in each panel is the same as Fig.~5 of \citet{bontemps96}.
}
\label{fig:14}
\end{figure}

The correlations between the observed outflow momentum flux and 
protostellar luminosity, or envelope mass, should suggest 
an underlying relationship between stellar outflow activity
and protostellar evolution.
In this section, we examine similar correlations from our simulations and compare them with the observations.
The outflow momentum flux for each model is plotted against the bolometric luminosity in Figure~\ref{fig:14} and against the envelope mass in Figure~\ref{fig:15}.
To derive the outflow momentum $F$,  we calculated total outflow momentum $(M V)_{\rm out}$ in the outflowing region ($v_r > c_{s,0}$) and divided that value by the elapsed time $t_{\rm out}$ after the outflow appearance 
\begin{equation}
F = \frac{(M V)_{\rm out}}{t_{\rm out} } = \frac{\int \rho\,v_{r, {\rm out}} \, dv }{ t_{\rm out}}.
\end{equation}
In the figures, the upper panels present the evolutionary tracks
for the models, and the lower panels show images every
thousand years.
Models with the different initial magnetic fields
and different rotation rates are separately plotted
in the left- and right-hand panels.
The standard model 3 is plotted in all panels for comparison.


In Figure~\ref{fig:14}{\it a}, the tracks first move from 
the left to the lower right, then turn around to the lower left.
These evolutionary tracks agree well with those expected in
\citet[][Fig.5]{bontemps96}.
For models plotted in Figure~\ref{fig:14}{\it a}, the outflows have momentum fluxes (i.e. the vertical axes) in the range of $10^{-5} < F/(\mv\,{\rm yr}^{-1}) < 10^{-4}$ at the emergence time of outflow, which gradually decrease until the outflow disappears.
On the contrary, the protostellar luminosities (i.e. the horizontal axes) increase after protostar formation and begin to decrease after reaching peak values of $L_{\rm ps}\sim10\lsun$  approximately $10^4-10^5$\,yr after protostar formation, as shown in Figure~\ref{fig:11}.
Reflecting both evolution of the outflow momentum and protostellar luminosity, the momentum flux evolves along the arrows in Figure~\ref{fig:14}{\it a}.


In Figure~\ref{fig:14}{\it b}, the evolutionary track of model 8 
qualitatively differs from that shown in Figure~\ref{fig:14}{\it a}. 
For this model, the angular momentum is mainly transferred by gravitational torque (Fig.~\ref{fig:12}), and a highly time-variable accretion is realized (\S\ref{sec:mass-accretion-rate} and Fig.~\ref{fig:11}).
Thus, with a time-variable accretion, the protostellar luminosity also shows a high time variability.
As a result, model 8 shows a zigzag evolutionary track of outflow momentum flux.
In addition, compared with the models in Figure~\ref{fig:14}{\it a}, the evolutionary track in model 6 (and 7) differs 
relatively. 
The outflow appears long after the protostar formation in model 6 (and 7), while it appears before protostar formation in other models (Table~\ref{table:3}).
For model 6, the outflow appears $1.316\times10^4$\,yr after the protostar formation.
By this epoch, the protostar sufficiently evolves and has a luminosity of $\sim30\lsun$.
Thus, in the diagram, the outflow momentum flux suddenly appears at ($L_{\rm bol}$, $F$) = ($\sim$30$\lsun$, $\sim$6$\times10^{-6}\,\mv\,{\rm yr}^{-1}$).
Then, the momentum flux moves only to the lower left because the
protostellar luminosity has already passed the peak.


Next, we focus on the parameter dependence of the evolutionary track.
The upper panels in Figure~\ref{fig:14} indicate that the evolutionary track for models with stronger magnetic field tends to lie in the lower right area.
This occurs because models with stronger magnetic fields have smaller outflow momentum (Fig.~\ref{fig:10}) but higher protostellar luminosity (Fig.~\ref{fig:13}), as described in \S\ref{sec:parameters}. 
In addition, because models with smaller rotation rates have a smaller outflow momentum and  higher protostellar luminosity (\S\ref{sec:parameters}), their evolutional track lie in the lower right area.

In Figure~\ref{fig:14}, with lower panels one can roughly compare observations with calculation results.
Since the momentum fluxes every thousand year are plotted in the panels, it is expected that  protostars are frequently observed in areas with densely grouped circles, while they are rarely observed in  sparsely-grouped areas.
With observed Class I sources, \citet{bontemps96} derived the best fit for linear correlation between the outflow momentum flux $F$ and protostellar bolometric luminosity  $L_{\rm bol}$ as
\begin{equation}
{\rm log} \ (F/\mv\, {\rm yr}^{-1} ) = -5.6 + 0.9\ {\rm log}\ (L_{\rm bol}/\lsun).
\label{eq:f-lbol}
\end{equation}
In Figure~\ref{fig:14}, we plotted the same correlation line (eq.~\ref{eq:f-lbol}) as shown in Fig.~5 of \citet{bontemps96}.
In the figure, Class 0 protostars are distributed in the upper left area towards the correlation line (solid line), while Class I (and II) protostars are roughly distributed along the correlation line.
Several Class I protostars are distributed in the lower left or lower right area towards the line.
Because the lifetime of Class I protostars is longer than that of Class 0 protostars,  the number of Class I protostars plotted in the figure is greater than that of Class 0 protostars.
The horizontal dispersion (or luminosity dispersion) is caused by the different evolutionary stages of the protostar (or different luminosities), while the vertical dispersion (or momentum flux), is likely attributed to different cloud parameters.
Protostars formed in clouds with stronger magnetic fields or slower rotations are distributed in the lower right area.
Thus, cloud parameters can be expected from observations of outflow momentum and protostellar luminosity.

In the figure, some Class 0 protostars are distributed in the (upper) left area, which indicates that the protostellar outflow emerges in a very early phase of Class 0 stage,  or prior to protostar formation.
When  protostellar outflow appears in the later evolutionary stage, protostars are distributed only in the lower right area towards the correlation line, as shown by model 6.
A comparison of the evolutionary tracks in the upper panels with open circles in lower panels reveals that no circle is plotted in the region of $L_{\rm bol} \lesssim 0.3\lsun$ and $F \gtrsim 5\times10^{-5}\mv\, {\rm yr}^{-1}$ because such protostars have an age of $t_{\rm ps}< 1000$\,yr.
If a protostar located in such an area is observationally
confirmed, it can be identified as a very young object ($<1000$\,yr), 
through which the very early stage of star formation can be investigated.

\subsection{Momentum Flux vs. Envelope Mass}
\begin{figure}
\includegraphics[width=150mm]{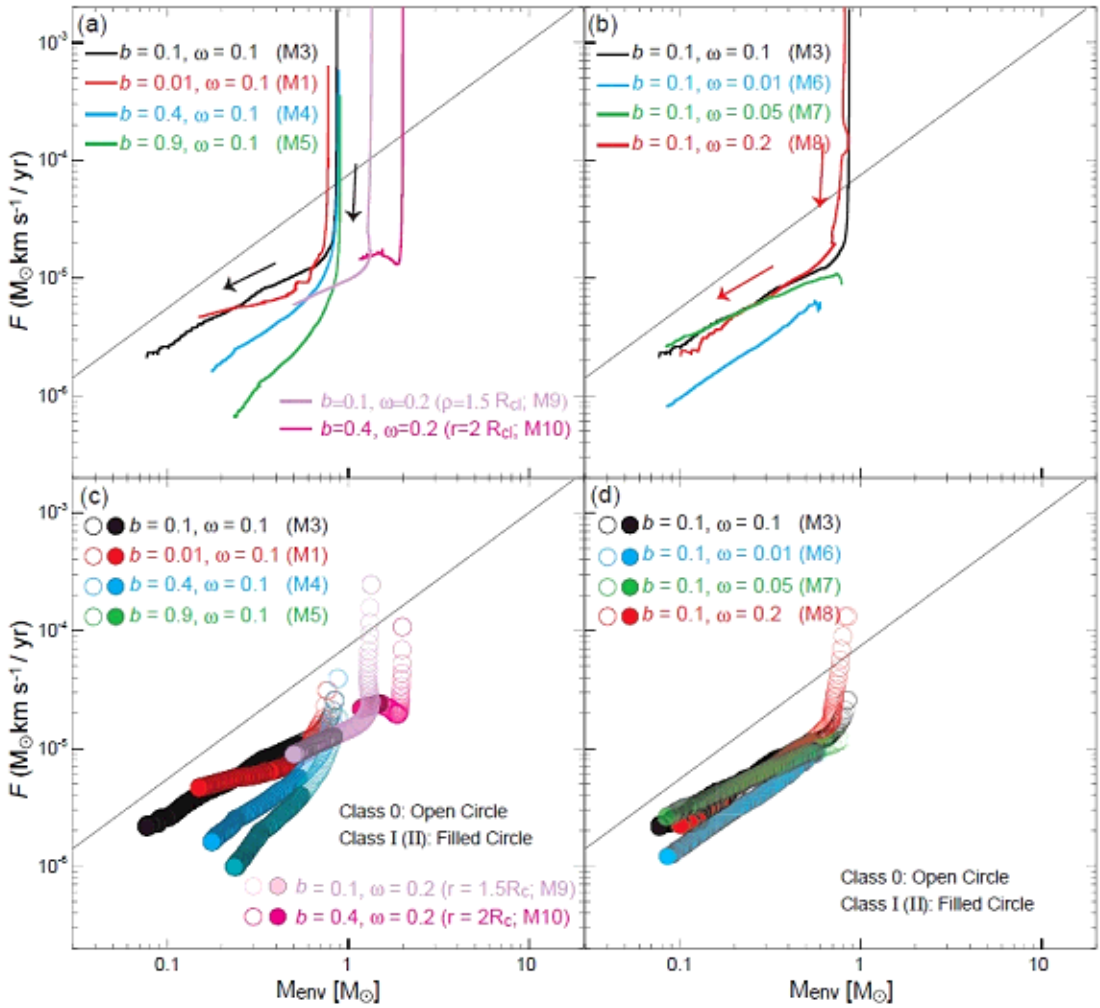}
\caption{
Same as in Fig.~\ref{fig:14} but against the envelope mass.
}
\label{fig:15}
\end{figure}

Figure~\ref{fig:15} shows the momentum flux against the envelope mass, in which the same correlation line is plotted as that in  Fig.~6 of \citet{bontemps96}, 
\begin{equation}
{\rm log} \ (F/\mv) = -4.15 + 1.1\ {\rm log}\ (M_{\rm env}/\msun).
\label{eq:f-menv}
\end{equation}
In the figure, the evolutionary track for each model moves 
from near the upper boundary to the lower left.
In all models except models 6 and 7, outflow appears prior to  protostar formation.
The envelope mass rarely decreases prior to protostar formation because the protostar cannot gain the mass of the infalling envelope.
Even before protostar formation, the infalling envelope slightly decreases because  part of the gas ($\sim0.01\,\msun$) has accreted onto the first core (i.e. the rotating disk). 
In contrast, the outflow momentum flux continues to decrease from its
emergence (Figure~\ref{fig:15}).
Therefore, prior to protostar formation, the momentum flux moves vertically downward nearly maintaining the initial envelope mass.
Then, following protostar formation, the envelope mass rapidly decreases (Fig.~\ref{fig:8}), and the evolutionary track moves to the lower left.
On the contrary, the evolutionary tracks for models 6 and 7 start from different points (near the center of the figure) in Figure~\ref{fig:15}{\it b} because the outflow appears long after the protostar formation during which time the envelope mass continues to decrease.

Models with initially different cloud masses (models 9 and 10) are also plotted in the left panels in Figure~\ref{fig:15}. 
Because the host cloud for these models is more massive than that in other models, the evolutionary tracks start from $M_{\rm env}\simeq 1.6\msun$ and 2.1$\msun$, respectively (Table~\ref{table:3}).
However, the evolutionary tracks for these models have the same trend as that in other models.
Thus, different cloud mass produces no qualitative differences in evolutionary tracks. 

The outflow momentum fluxes at every thousand years plotted in the lower panels of the figures are mainly distributed near the correlation line, which indicates that the outflow momentum fluxes derived in this calculation agree with  the observation of \citet{bontemps96}.
Moreover, we compared these panels with the observations of \citet[][Fig.~4]{hatchell07} and \citet[][Fig.~6]{curtis10}, and confirmed good agreement between our results and observations.
In addition, Figure~\ref{fig:15} shows that an outflow with $F\gtrsim 3\times 10^{-4}\,\mv\,{\rm yr}^{-1}$ indicates a very early phase of the outflow (before protostar formation).

\subsection{Evolutionary Relation between Outflow and Cloud Core}
With a substantial amount of observational data, \citet{bontemps96} expected  that the decrease in outflow momentum flux and envelope mass is an evolutionary effect independent of protostellar luminosity, or protostellar mass.
They also demonstrated that the outflow momentum flux ($F$) and envelope mass ($M_{\rm env}$) are proportional to the bolometric luminosity as 
$F \propto L_{\rm bol}^{1.04 \pm 0.2}$ and $M_{\rm env} \propto L_{\rm bol}^{0.56\pm0.2}$, respectively.
Then, to remove any luminosity dependence, they created a diagram
(\citealt{bontemps96} Fig.~7), in which the outflow efficiency ($F
\cdot c/ L_{\rm bol}$; dimensionless, where $c$ is the speed of light)
is plotted against $M_{\rm env}/L_{\rm bol}^{0.6}$.
In their diagram, outflow momentum fluxes for Class 0 and I protostars are separately distributed.
Class 0 protostars are widely distributed, while Class I protostars are clustered in a narrow area.
They concluded that outflow momentum flux depends on only one basic parameter related to $M_{\rm env}$ rather than separately on age and stellar mass.

\begin{figure}
\includegraphics[width=150mm]{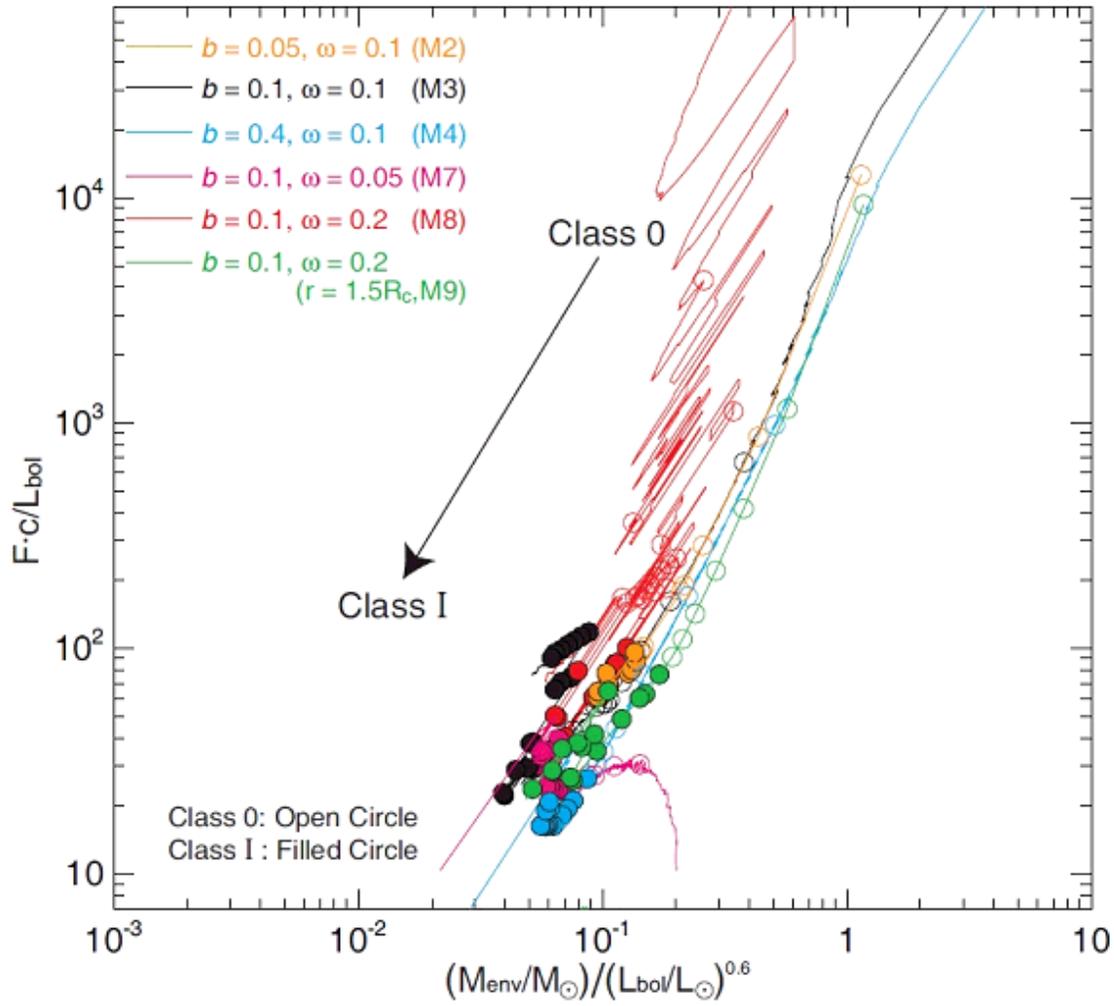}
\caption{
Momentum flux divided by the protostellar luminosity against the ($M_{\rm env}/\msun$)/$(L_{\rm bol}/L_{\rm sun})^{0.6}$.
}
\label{fig:16}
\end{figure}

Our results support their conclusion.
To compare observations with our results in more detail,  we used our calculation data to create the same diagram (Fig.~\ref{fig:16}) as that in Fig. 7 of \citet{bontemps96}, in which $F\cdot c/L_{\rm bol}\, (\equiv y_{\rm B})$ is plotted against $(M_{\rm env}/L_{\rm bol})^{0.6}\, (\equiv x_{\rm B})$ for models 2, 3, 4, 7, 8 and 9.
As shown in \citet{bontemps96}, the Class 0 and I protostars are separately distributed in Figure~\ref{fig:16}.
Class I protostars are clustered in a narrow area of $0.03<x_{\rm B}<0.2$ and $10<y_{\rm B}<120$ for our results and appeared in an area of $0.01<x_{\rm B}<0.2$ and $10<y_{\rm B}<600$ for \citet{bontemps96}.
In addition, Class 0 protostars are widely distributed in the range of $x_{\rm B}>0.07$ and $30<y_{\rm B}< 10^4$ for our results, while \citet{bontemps96} showed results of  $x_{\rm B}>0.2$ and $200<y_{\rm B}<10^4$.
The different distribution between Class 0 and I protostars is caused by different conditions of outflow, or different evolutionary stages.
As shown in Figure~\ref{fig:10}, the outflow momentum inside the host cloud begins to decrease by the end of the Class 0 stage.
Thus, during the Class I stage, no powerful outflow is driven by the circumstellar disk.
That is, the protostellar outflow cannot gain additional momentum during the Class I stage.
However, the outflow momentum is nearly conserved and the outflow propagates into the interstellar space keeping the momentum acquired during the Class 0 stage.
It should be noted that the outflow momentum gradually decreases over a lengthy period because it interacts with the interstellar medium.
Therefore, the outflow momentum hardly changes during the Class I stage, or the momentum driven phase.
In addition, the main accretion phase has already ended by the time of the Class I stage, as shown in Figure~\ref{fig:11}.
Thus, the protostellar luminosity is mainly supplied by the Kelvin-Helmholtz contraction rather than by release of accretion energy.
Because the Kelvin-Helmholtz timescale is as long as $\sim10^6$\,yr just
after the mass accretion ceases, protostellar luminosity change is minimal during this phase.
As a result, the evolutionary track during the Class I stage rarely moves in the vertical direction because both $F$ and $L_{\rm bol}$ are hardly changed.
In addition, the protostellar luminosity is roughly proportional to $M_{\rm env} \propto L_{\rm bol}^{0.6}$ (e.g. $L_{\rm bol} \propto M_{\rm env}^{1.67}$) during the Class I stage, as described in the upper panel in Figure~\ref{fig:13}.
Thus, the evolutionary track also rarely moves in the horizontal direction.
As a result, the evolutionary track remains in a small area during the Class I stage.

On the other hand, just after protostar formation, both the outflow momentum (Fig.~\ref{fig:10}) and luminosity (Fig.~\ref{fig:13}) increase.
Next, during the Class 0 stage,  protostellar luminosity turns to decrease after the outflow momentum turns to decrease.
The infalling gas first accretes onto the circumstellar disk.
Part of the accreted gas in the circumstellar disk is blown away by the protostellar outflow, which is powered by the accretion from the infalling envelope.
Thus, as the gas accretion onto the circumstellar disk weakens or the infalling envelope is depleted, the outflow weakens, and its momentum inside the host cloud begins to decrease.
On the contrary, protostellar luminosity is mainly related to gas accretion from the circumstellar disk.
The accreted gas from the infalling envelope remains in the circumstellar disk for a short period.
Thus, protostellar luminosity does not decrease just after depletion of the infalling envelope because  protostar luminosity is caused by accretion from the circumstellar disk.
Therefore, near the end of the Class 0 stage, the outflow momentum rarely increases (or the outflow momentum flux $F$ begins to decrease), then the protostellar luminosity $L$ decreases some time later. 
As a result, the evolutionary track of $F/L_{\rm bol}$ moves downward in Figure~\ref{fig:16}, because the numerator first decreases.
In addition, at this stage, protostellar luminosity decreases, or increases, minimally while the infalling envelope rapidly decreases, as shown in Figure~\ref{fig:13}({\it a}).
Thus, the ratio of the envelope mass to protostellar luminosity $M_{\rm env}/L_{\rm bol}$ tends to decrease, and the evolutionary track moves towards the left.
Therefore, during the Class 0 stage, the evolutionary track moves to the lower left as shown in Figure~\ref{fig:16}.
It should be noted that the evolutionary track does not move widely left because the
protostellar luminosity by the Kelvin-Helmholtz contraction dominates
that by accretion in the later Class I stage and the protostar is less dark.


In summary, reflecting the rapid attenuation of outflow (momentum flux) during the Class 0 stage, the evolutionary track moves in a wide range, and the protostar is widely distributed, as shown in Figure~\ref{fig:16}.
During the Class I stage, the outflow is in a momentum driven, or snow-plough, phase without additional momentum.
Thus, the evolutionary track remains in a small area.
Therefore, the different distribution of protostars on the diagram between Class 0 and I protostars is caused by the outflow condition such that outflow is powered by mass accretion during the Class 0 stage and is in a momentum driven phase during the Class I stage.

\section{Discussion}
\subsection{Dependence of Host Cloud Mass}
\label{sec:host-cloud-mass}
In this study, we investigated the cloud evolution with different cloud parameters of the magnetic field $B_0$, rotation rate $\Omega_0$ and initial cloud mass $M_{\rm cl}$.
As listed in Table~\ref{table:1}, we adopted a wide range of magnetic field strength ($ 7.8 \times 10^{-6}\, \mu {\rm G} < B_0 < 7.4 \times 10^{-5}\, \mu {\rm G}$) and rotation rate ($1.0\times10^{-14}\, {\rm s}^{-1}< \Omega_0 < 2.1\times10^{-13}\, {\rm s}^{-1}$), while we only adopted three different initial cloud masses of $M_{\rm cl}=1.05\msun$, $1.6\msun$ and $2.1\msun$. 
The purpose of this study is to associate the properties of the protostellar outflow with the envelope mass.
Thus, we may have to investigate the cloud evolution with a more wide range of initial cloud mass.
However, with an initially massive cloud core, we cannot calculate the cloud evolution until the infalling envelope is depleted because it needs a huge computational time.
In this study, we could calculate the cloud evolution until the mass of the infalling envelope decreases to $\lesssim 10$\% of the initial cloud mass for models with $M_{\rm cl}=1.05\msun$ (models 1-8), while $52\%$ for model 10 which has the  host cloud mass of $M_{\rm cl}=2.1\msun$.
Thus, it is considerably difficult to calculate the cloud evolution until Class I and II stages with a massive host cloud.
However, as a result of the calculation, formed protostars have a mass of $0.24-0.71\msun$, as summarized in Table~\ref{table:2}.
The initial mass function seems to have peak at $\lesssim 1\msun$ \citep[e.g.][]{kroupa01}, and the cloud mass function in various star forming regions has a peak around $1\msun$ \citep[e.g.][]{andre10}.
Thus, we believe that we investigated the evolution of most typical clouds observed in star forming regions, or the evolution of typical stars, in this study.
On the contrary, a massive protostar formed in a massive cloud ($\gg 1\msun$) shows a more massive, or larger momentum, outflow.
Although such massive outflows are small in number, they are considered to be preferentially observed.
In this subsection, we discuss the dependence of the initial cloud mass on the protostellar outflow.

\subsubsection{Outflow Energy}
\begin{figure}
\includegraphics[width=150mm]{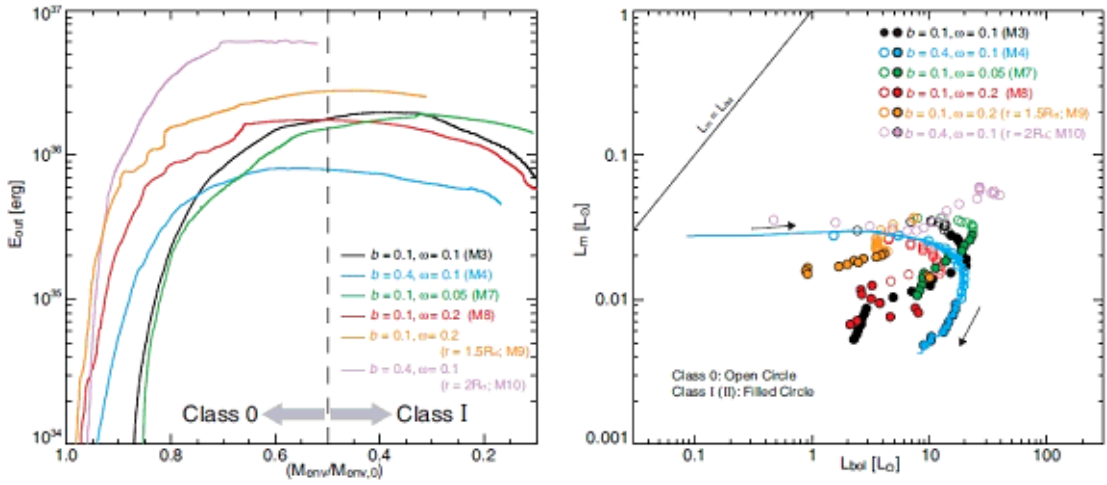}
\caption{
Outflow energy $E_{\rm out}$ against the normalized envelope mass (left) and outflow kinematic luminosity $L_{\rm m}$ against the protostellar bolometric luminosity $L_{\rm bol}$ (right) for models 3, 4, 7, 8, 9 and 10.
The relation $L_{\rm m}=L_{\rm bol}$ is plotted in the right panel. 
The evolutionary track for model 4 is also plotted by the blue solid line in the right panel.
}
\label{fig:17}
\end{figure}

In the left panel in Figure~\ref{fig:17} , the outflow energies $E_{\rm out}$, which is defined as
\begin{equation}
E_{\rm out} = \dfrac{1}{2} \int  \rho_{\rm out}\, v_{\rm out}^2 \, dv,
\end{equation}
where $\rho_{\rm out}$ and $v_{\rm out}$ are the density and velocity of outflow at each point, respectively, 
is plotted against the normalized envelope mass for models 3, 4, 7, 8, 9 and 10.
The outflow energy has a peak around $\sim10^{36}-10^{37}$\,erg, and begins to decrease during the Class I stage.
The outflow energies derived in our calculation are comparable to the
observational estimates.
\citet{curtis10} observed 45 outflows in the Perseus molecular clouds and showed that the outflow energies lie in the range of $10^{35}\,{\rm erg} <E_{\rm out}<10^{37}\,{\rm erg}$ except for a highly powerful anomaly.
On the contrary, a massive protostar seems to drive a more powerful outflow \citep{wu04}.
Recently, \citet{motogi11} showed that the outflow around a young massive protostar that is embedded in $\sim 200\msun$ envelope has the kinetic energy of $>10^{46}$\,erg.

The left panel in Figure~\ref{fig:17} indicates that the outflow kinetic energy strongly depends on the initial cloud mass, while it weakly depends on the cloud parameters of the magnetic field and rotation rate.
The outflow kinematic energy for model 10 is about ten times larger than that for model 4 that has the same cloud parameters of magnetic field ($b$) and rotation ($\omega$) as in model 10 but the different host cloud mass ($M_{\rm cl}=2.1\msun$ for model 10, and  $1.1\msun$ for model 4).
The outflow kinetic energy (and momentum) continues to increases until the infalling envelope almost halves (e.g. during the Class I stage: Fig.~\ref{fig:10}).
A massive cloud has a longer duration of the Class 0 stage during which a sufficient massive infalling envelope can give  power to drive the outflow.
 Note that the outflow energy largely dissipates inside the host cloud  by the interaction between outflow and infalling envelope, in which the outflow loses its energy radiatively  at the shock.
Note also that since the mass of the infalling envelope depends on the initial cloud mass, outflows are expected to have different energies among models with different cloud mass. 
As a result, the protostellar system formed in a more massive envelope can drive the protostellar outflow for a longer duration, and outflow in such a system can acquire a larger kinematic energy, or momentum.
Thus, it is expected that outflow with a larger kinematic energy tends to be observed in a massive infalling envelope.
However, it is also expected that such powerful outflows are not ubiquitous due to the rarity of a massive cloud core ($M_{\rm cl}\gg 1\msun$).

\subsubsection{Outflow Kinematic Luminosity}
In the right panel in Figure~\ref{fig:17}, the outflow kinematic luminosity $L_{\rm m}$, which is defined as 
\begin{equation}
L_{\rm m} = \dfrac{E_{\rm out}}{t_{\rm out}},
\end{equation}
is plotted against the protostellar luminosity for every $3000$\,yr after the protostar formation.
The evolutionary track for model 4 (the blue solid line) is also plotted. 
In the figure, protostars  are distributed in the range of $0.5\lsun<L_{\rm bol}<50\lsun$ and $0.005\lsun<L_{\rm m}<0.07\lsun$.
In the same range of the bolometric luminosity, almost the same distribution of protostars is seen in the observations \citep{cabrit92, wu04}.

Both our results and  observations indicate that the mechanical luminosity weakly depends on the protostellar bolometric luminosity.
In the panel, the evolutionary track of $L_{\rm m}$ is much shallower than  $L_{\rm m}\propto L_{\rm bol}$; it moves horizontally during the main accretion phase and moves to the lower left after the main accretion phase.
This means that the outflow driving is not directly related  to protostellar properties such as luminosity, mass and age.
The protostellar properties are determined by the accretion process onto the protostar from the circumstellar disk.
On the other hand, outflow properties are determined by the accretion process onto the circumstellar disk from the infalling envelope.
Thus, the protostellar outflow is related to the envelope mass, while the protostellar luminosity is related to the circumstellar disk mass.
Therefore, as pointed out by \citet{bontemps96}, it seems that the protostellar outflow is not primary related to the protostellar luminosity.
This implies that the protostellar outflow is not powered by the
radiation of the protostar.
In addition, there is no significant difference in the peak value of the mechanical luminosity among models with different host cloud masses.
This indicates that the outflow is powered by the accretion at a constant rate because the mechanical luminosity, which is the outflow energy divided by its lifetime, is almost constant during the main accretion phase.

In summary, the acquisition rate of the outflow energy or momentum is independent of the initial cloud mass, while it slightly depends on the cloud parameters of the magnetic field and rotation rate.
The protostellar outflow gains its energy or momentum at a constant rate during the main accretion phase.
Since a massive cloud has a longer period of the main accretion phase that almost corresponds to duration of the Class 0 stage, the outflow appeared in a massive cloud core can possess  a larger energy and momentum.
In other words, a massive star shows a more powerful outflow  because a relatively massive star forms in a massive cloud.
Therefore, the outflow (peak) energy, or momentum, reflects its host cloud mass.

\subsection{Cloud Parameters}

In \S\ref{sec:observations}, we have shown that the outflow in the Class 0
stage is essentially different from that in the Class I stage
(see also \citealt{bontemps96}).
The circumstellar disk forcefully drives outflow during the Class 0
stage, while the driving force from the circumstellar disk ceases and
outflow is in the momentum-driven (slow-plough) phase during the Class I
stage.
As a result, Class 0 and I protostars are distributed in different
regions in Figure~\ref{fig:14}, in which the outflow momentum flux is
plotted against the protostellar bolometric luminosity.


Observations often show a large scatter of the outflow momentum flux
even among protostars which are in
the same evolutionary stage \citep{hatchell07}.
Such scatter may be attributed to variations of the cloud parameters such as  
the strength of the magnetic field and rotation rate.
In Figure~\ref{fig:14}, the protostar formed in a cloud with
strong magnetic field (e.g. model 5) or slower rotation (model 6 and 7)
has a relatively smaller momentum flux.
The outflow momentum in these models is about one order of magnitude
smaller than the other models. 
A scatter of the mechanical luminosity is also seen in the right panel in 
Figure~\ref{fig:17}.
Thus, it is natural that the observation shows a scatter of outflow
properties because the host cloud has different cloud parameters
\citep{caselli02,crutcher99}.

\subsection{Protostellar Mass and Star Formation Efficiency}

The protostellar outflow reduces the star formation efficiency 
because it ejects materials from the host cloud 
into the interstellar space.
The magnetic field and rotation drive the outflow  \citep{blandford82,uchida85}.
Thus, it is expected that both magnetic field and cloud rotation control the outflow and star formation efficiency.
As summarized in Table~\ref{table:2}, even in clouds having the same mass, different magnetic field strengths and rotation rates bring different results.
For example, a half of the cloud mass is ejected by the protostellar outflow for model 2, while only 10\% of the host cloud mass is ejected for models 5 and 6, in which the magnetic field strength for models 5 and 6 is stronger than that for model 2.
As described in \S\ref{sec:parameters}, both a moderate strength of the magnetic field and rapid rotation rate are necessary to drive a powerful outflow.
Neither weak magnetic field nor slow rotation can drive a powerful outflow.
In addition, in a strongly magnetized cloud, the magnetic braking
effectively transfers the disk angular momentum and delays the
formation of a sufficiently large disk that is the driver of the outflow
\citep{machida11b}.
Thus, too strong magnetic field also suppresses a powerful outflow driving.

In models with the same cloud mass of $M_{\rm cl}=1.05\msun$ (models 1-8), the protostars have their mass in the range of  $M_{\rm ps}=0.24-0.71\msun$, which corresponds to the star formation efficiency of $\varepsilon = 0.23-0.68$.
As seen in Table~\ref{table:2}, the star formation efficiency is high in a cloud with stronger magnetic field or slower rotation.
In such a cloud, a relatively small circumstellar disk drives a relatively weak outflow that cannot eject a large fraction of cloud mass from the host cloud, and a relatively massive protostar finally forms.
Thus,  the outflow efficiency is inversely correlated with the star formation efficiency.

The outflow efficiency is also inversely correlated with the protostellar luminosity in the main accretion phase.
As described in Table~\ref{table:2}, at the end of the Class 0 stage,
the more luminous protostar has the less massive outflow (models 5, 6
and 7) and vice versa (models 1,2 and 3).
Figure~\ref{fig:14} also shows that the luminous protostars have smaller
outflow momenta.
The protostellar luminosity is related to the outflow efficiency through the efficiency of the angular momentum transfer in the circumstellar disk.
In a strongly magnetized cloud, the disk angular momentum is effectively transferred not by the outflow but by the magnetic braking that promotes the mass accretion from the circumstellar disk onto the protostar without mass ejection, and a relatively massive protostar forms.
Since the accretion luminosity is proportional to both the accretion
rate and protostellar mass, both the higher mass accretion and massive
central object makes the protostar luminous. 

On the contrary, with a relatively weak magnetic field, the angular
momentum transfer is inefficient with the weak magnetic braking.
In such clouds, the strong disk rotation drives the powerful outflow
suppressing the rapid (or effective) mass accretion onto the protostar.
Therefore, the less luminous protostar  tends to have a more powerful
outflow.


Note that this trend may not be applicable among clouds with initially different masses because a massive protostar can form with a powerful outflow when the initial cloud has a sufficient mass, irrespective of the magnetic field strength and rotation rate.
However, we can observe these trends in some star forming regions where low-mass stars born in clouds with similar mass.
We need a more detailed observation of the core mass function in various star forming regions to confirm the relation between the protostellar luminosity and outflow properties.

\subsection{Molecular Outflow Model}
\label{sec:outflowmodel}
A very young protostar sometimes shows two, or more, different types of flow: high-velocity (or optical) jet and low-velocity (or molecular) outflow.
The molecular outflow is frequently observed around a protostar \citep{wu04,zhang05,hatchell07}, while the optical (or the high-velocity molecule) jet is occasionally observed in molecular outflow \citep{mundt83,richer92,mitchell94,arce02,velusamy07}. 
Observations indicate that a high-velocity jet is enclosed by the low-velocity outflow.
Since molecular outflow has a large amount of mass \citep[e.g.][]{downes03,stoji06}, it affects the star formation efficiency and greatly contributes to the star formation process \citep{nakano95}.
However, in observations, it is difficult to specify the driving mechanism for molecular outflow because we cannot directly observe the driving region that is embedded in a dense cloud core.
Thus, many authors have tried to theoretically clarify it.

Historically, the entrainment mechanism has been proposed to explain the molecular outflow driving, in which one imagined the low-velocity outflow entrained by a high-velocity jet.
In other words, the primary jet injects its momentum into the surrounding gas resulting in molecular outflow.
This mechanism was simply formulated with observations of  high-velocity well-collimated jets embedded in the wide-angle low-velocity outflow \citep{mundt83}.
There are a variety of entrainment models such as the wind-driven shell \citep{shu91,matzner99}, jet-driven bow shock \citep{raga93a,raga93b}, and jet-driven turbulent \citep{cant91,lizano95} models.
It seems that some models are possible to explain a number of observations of molecular outflow and some is difficult to explain general feature of outflows \citep{cabrit97,lee00,arce07}.
Around the protostar, the molecular outflows are frequently observed, while the high-velocity component is rarely observed.
Thus, we have to posit the undetected jet to the model molecular outflow with the entrainment mechanism, although the high-velocity jet may be invisible by chance.
In addition, to explaining observations, one can adjust jet and outflow models by changing a number of artificial, or ambiguous, parameters such as the jet speed and momentum conversion efficiency between jet and outflow, environment density, strength of magnetic field, etc.

Aside from the entrainment model, there is an entirely-different approach for theoretically investigating the molecular outflow.
\cite{tomisaka02} calculated the evolution of the whole molecular cloud without any artificial setting and reproduced the low-velocity outflow in the collapsing cloud \citep[see also][]{tomisaka98,tomisaka00}.
In his study, the first core forms before the protostar formation and  directly drives the low-velocity outflow.
He also reproduced the high-velocity jet which is driven near the protostar.
Then, both low- and high-velocity flows appeared in the collapsing cloud have been confirmed by many authors \citep[e.g.][]{machida04,matsu04,banerjee06,hennebelle08a,commerson10,tomida10a,burzel11,seifried12,price12}.
The high-velocity jet and low-velocity outflow are naturally reproduced without any artificial setting in cloud collapse calculations, while the jet is artificially input to entrain the molecular outflow in entrainment models. 
However, in the cloud collapse calculations, the authors could not investigate a long-term evolution of jet and outflow because they had to resolve the protostar itself that is the driver of the jet \citep{banerjee06,machida08b,tomida12}.

In this study, at the expense of spatial resolution around the protostar, we could investigate the evolution of molecular outflow and protostellar system for $\sim10^5$\,yr with sink treatment.
However, no high-velocity jet appears because we did not resolve the region in the proximity of the protostar ($r<1$\,AU) where the high-velocity jet is driven \citep{machida08b}. 
Thus, we could not investigate the effect of the jet on the molecular outflow.
On the contrary, since we resolve the magnetically inactive region inside the circumstellar disk \citep{machida07,tomida12}, we could precisely investigate the low-velocity outflow driven by the circumstellar disk.
Note that the low-velocity outflow is launched outside of the
magnetically inactive region \citep{inutsuka10,inutsuka12}.


In \S\ref{sec:results} and \ref{sec:observations}, we have shown that
our results agree well with observations of molecular outflows; the
outflow momentum (flux) and energy are comparable to the observations.
Since we have not focused on the effect of a high-velocity jet, 
we do not exclude the entrainment mechanism.
In reality, a part of the infalling gas might be entrained by the high-velocity jet.
However, our results suggest that the observed molecular outflows can be
explained only by the flow directly driven by the circumstellar disk.
In contrast to the entrainment models, 
the direct cloud collapse simulations explain the molecular outflow
well without any fine-tuning of model parameters.
Although further simulations with high spatial resolutions would be useful to
examine the effect of the high-speed jet, our current simulations mostly
explain the available observations of the low-velocity outflow.


\section{Summary}

In this study, we calculated the cloud evolution from the
pre-stellar core stage until almost all the envelope gas dissipates by
the protostellar outflow.
In the collapsing cloud, the first core forms prior to the protostar
formation and evolves into the circumstellar disk.
The low-velocity outflow is first driven by the first core, 
then driven from the outer part of the disk which is magnetically
active. 
The outflow driving region extends as the circumstellar disk grows.
Before the outflow breaks out of the cloud, the outflow propagates along
hourglass-shaped magnetic field lines that open up with increasing
the distance from the equator.
As a result, the outflow has a wide-opening angle in a cloud scale.
After the outflow penetrates the cloud, it propagates along the
interstellar magnetic field lines.
The head of the outflow travels over
$\sim10^5$\,AU in $10^5$\,yr.
In contrast, the horizontal extension of the outflow is limited
by the cloud scale.
The width of the outflow reflects its host cloud size.
The outflow thus extends only in the vertical direction, 
and its collimation gets improved in this stage.


The properties of the calculated outflows such as 
the outflow momentum, energy and mass
agree well with those of observed outflows.
Our simulations also explain the physical structure of the observed
outflows.
These support the picture that the low-velocity molecular outflows
which are frequently observed around protostars are directly
driven in the circumstellar disk, or the first core.
The entrained flow is not necessary to explain observations,
though some amount of gas might be entrained by the high-velocity
component.


Our calculations show the same correlations between the
outflow momentum flux, protostellar luminosity and envelope mass
as in observations.
These correlations differ between Class 0 and I protostars, 
which is explained with the different evolutionary stages 
of low-velocity outflow.
In the Class 0 stage, the sufficient gas accretes onto the circumstellar
disk, and the outflow powered by the accretion is driven from the disk.
The outflow momentum is continuously supplied from the disk, or
the accreting gas, during this stage. 
However, the outflow gradually ceases 
as the infalling envelope gets depleted.
In the Class I stage, the outflow hardly acquires its momentum from 
the accreting gas and enters into the momentum-driven
snow-plough phase.


The protostellar outflow ejects  half of the cloud mass from the
host cloud and limits the star formation efficiency down to $\sim 50\%$,
whose exact value depends on the cloud parameters 
such as the magnetic field strength and rotation rate.
A stronger magnetic field excessively transfers 
the angular momentum and forms a relatively small circumstellar disk 
that drives a relatively weak outflow.
A weak magnetic field and slow cloud rotation also weaken 
the outflow driving force. 
On the contrary, 
the clouds with moderate magnetic fields and rotation,
$b_0=0.05-0.4$ ($\mu\simeq3-10$) and
$\beta_0=0.01-0.04$, show considerably powerful outflows.
This parameter range agrees with the observational estimates
\citep[e.g.][]{crutcher99,caselli02}.
Therefore, the protostellar outflow should determine the final stellar mass 
and significantly affect the early evolution of the low-mass protostars.

\section*{Acknowledgments}
We have benefited greatly from discussions with ~T. Nakano and ~K. Tomida. 
We are very grateful to an anonymous reviewer for a number of useful suggestions and comments.
Numerical computations were carried out on NEC SX-9 at Center for Computational Astrophysics, CfCA, of National Astronomical Observatory of Japan.

\clearpage
\begin{table}
\caption{Model parameters}
\label{table:1}
\begin{center}
\begin{tabular}{c|cccccc|ccccc} \hline
{\footnotesize Model} & 
$b$ & $\omega$ & $n_{\rm c,0}$ {\scriptsize [cm$^{-3}$]} &  $R_{\rm c}$ {\scriptsize [AU]} &  $M_{\rm cl}$ {\scriptsize [$\msun$]} 
& $B_0$ {\scriptsize [G]} &  $\Omega_0$ {\scriptsize [s$^{-1}$]} & $\alpha_0$ & $\beta_0$ & $\gamma_0$ & $\mu $ \\
\hline
1 & 0.01 & 0.1 & $ 6 \times10^5$  & $6.1\times10^3$ & 1.05  & $7.8\times10^{-6}$ & $1.0\times10^{-13}$ & 0.5 & 0.01 & 0.006 & 23 \\
2 & 0.05 & 0.1 & $ 6 \times10^5$  & $6.1\times10^3$ & 1.05  & $1.8\times10^{-5}$ & $1.0\times10^{-13}$ & 0.5 & 0.01 & 0.03 & 9.9 \\
3 & 0.1 & 0.1 & $ 6 \times10^5$  & $6.1\times10^3$ & 1.05  & $2.5\times10^{-5}$ & $1.0\times10^{-13}$ & 0.5 & 0.01  & 0.06 & 7.0 \\
4 & 0.4 & 0.1 & $ 6 \times10^5$  & $6.1\times10^3$ & 1.05  & $5.0\times10^{-5}$ & $1.0\times10^{-13}$ & 0.5 & 0.01 & 0.23 & 3.5 \\
5 & 0.9 & 0.1 & $ 6 \times10^5$  & $6.1\times10^3$ & 1.05  & $7.4\times10^{-5}$ & $1.0\times10^{-13}$ & 0.5 & 0.01 & 0.52 & 2.4 \\
6 & 0.1 & 0.01 & $ 6 \times10^5$  & $6.1\times10^3$ & 1.05  & $2.5\times10^{-5}$ & $1.0\times10^{-14}$ & 0.5 & 0.0001 & 0.06 & 7.0 \\
7 & 0.1 & 0.05 & $ 6 \times10^5$  & $6.1\times10^3$ & 1.05  & $2.5\times10^{-5}$ & $5.2\times10^{-14}$ & 0.5 & 0.002 & 0.06 & 7.0 \\
8 & 0.1 & 0.2 & $ 6 \times10^5$  & $6.1\times10^3$ & 1.05  & $2.5\times10^{-5}$ & $2.1\times10^{-13}$ & 0.5 & 0.039 & 0.06 & 7.0  \\
9 & 0.1 & 0.2 & $ 6 \times10^5$  & $9.2\times10^3$ & 1.6  & $2.5\times10^{-5}$ & $2.1\times10^{-13}$ & 0.44 & 0.070 & 0.11 & 4.7 \\
10 & 0.4 & 0.2 & $ 6 \times10^5$  & $1.2\times10^4$ & 2.1  & $5.0\times10^{-5}$ & $2.1\times10^{-13}$ & 0.41 & 0.11 & 0.76 & 1.8 \\
\hline
\end{tabular}
\end{center}
\end{table}

\begin{table}
\setlength{\tabcolsep}{2pt}
\caption{Calculation Results}
\label{table:2}
\begin{center}
\begin{tabular}{c|||ccccc||||||||ccccc||||||||cccccc} \hline 
  & \multicolumn{5}{|c|}{ Class 0 } & \multicolumn{5}{|c|}{ Class I } & \multicolumn{4}{|c|}{ Class II or E.O.C$^*$ } \\ \hline
 {\footnotesize Model}  & $M_{\rm ps,0}$ & $M_{\rm disk,0}$ & $M_{\rm out,0}$    & $L_{\rm ps,0}$ & $t_{\rm 0}$  
                        & $M_{\rm ps,I}$ & $M_{\rm disk,I}$ & $M_{\rm out,I}$ & $L_{\rm ps,I}$ & $t_{\rm I}$ 
                        & $M_{\rm ps,II}$ & $M_{\rm disk,II}$ & $M_{\rm out,II}$ & $M_{\rm env}$ \\ \hline
1    &0.16 &0.36 &0.12 &6.2 &2.4$\times$10$^4$        & ---& ---& ---& ---& ---                      &0.24 &0.28 &0.42  & 0.12 \\ 
2    &0.19 &0.31 &0.16 &7.9 &3.5$\times$10$^4$        &0.26 &0.30 &0.39 &1.4 &7.1$\times$10$^4$      &0.26  &0.27 &0.49  & 0.07  \\  
3    &0.26 &0.21 &0.14 &14.1 &2.7$\times$10$^4$        &0.47 &0.14 &0.34 &2.6 &9.2$\times$10$^4$     &0.47 &0.12 &0.38  & 0.07 \\ 
4    &0.31 &0.18 &0.08 &19.3 &3.4$\times$10$^4$        &--- &--- &--- &--- &---                      &0.50 &0.21 &0.21  & 0.16 \\
5    &0.32 &0.17 &0.03 &21.8 &3.7$\times$10$^4$        &--- &--- &--- &--- &---                      &0.54 &0.21 &0.09  & 0.20 \\ 
6    &0.33 &0.17 &0.01 &30.8 &2.0$\times$10$^4$        &0.69 &0.18 &0.08 &7.2 &8.2$\times$10$^4$    &0.71 &0.18 &0.09  & 0.05 \\ 
7    &0.31 &0.20 &0.08 &23.0 &2.6$\times$10$^4$        &0.52 &0.19 &0.23 &4.4 &7.4$\times$10$^4$    &0.52 &0.20 &0.27 &   0.05 \\ 
8    &0.25 &0.12 &0.31 &12.5 &3.8$\times$10$^4$        &0.33 &0.12 &0.5 &2.5 &1.1$\times$10$^5$   &0.34 &0.13 &0.48 &  0.09 \\ 
9    &0.23 &0.30 &0.46 &4.1 &8.9$\times$10$^4$        &--- &---&---&---&---                          &0.27 &0.37 &0.53 & 0.49\\ 
10   &---&---&---&---&---        &--- &--- &--- &--- &---                                            &0.49 &0.28 &0.30 & 1.10 \\ 
\hline
\end{tabular}
\end{center}
* E.O.C means the end of calculation.
\end{table}

\begin{table}
\caption{Protostar formation and Outflow Emergence epochs}
\label{table:3}
\begin{center}
\begin{tabular}{c|cccccc|ccccc} \hline
{\footnotesize Model} & 
$t_{\rm ps,0}$ [yr] & $t_{\rm out,0}$ [yr]  \\
\hline
1 &  $7.5714\times10^4$ & $7.2963\times10^4$  \\
2 &  $7.4285\times10^4$ & $7.0360\times10^4$  \\
3 &  $7.3127\times10^4$ & $7.0466\times10^4$  \\
4 &  $7.5860\times10^4$ & $7.1786\times10^4$  \\
5 &  $8.2225\times10^4$ & $7.8861\times10^4$  \\
6 &  $6.8280\times10^4$ & $8.1441\times10^4$  \\
7 &  $7.0053\times10^4$ & $7.3155\times10^4$  \\
8 &  $9.2215\times10^4$ & $7.5304\times10^4$  \\
9 &  $9.3039\times10^4$ & $7.5246\times10^4$  \\
10 & $8.8585\times10^4$ & $8.0162\times10^4$  \\
\hline
\end{tabular}
\end{center}
\end{table}


\clearpage

\clearpage

\clearpage

\clearpage

\clearpage

\clearpage

\clearpage

\clearpage

\clearpage

\clearpage

\clearpage

\clearpage

\clearpage

\clearpage

\clearpage

\clearpage

\clearpage


\begin{thebibliography}{}{}
\bibitem[Andre et al.(1993)]{andre93} 
Andre, P., Ward-Thompson, D., \& Barsony, M.\ 1993, ApJ, 406, 122 

\bibitem[Andre \& Montmerle(1994)]{andre94} 
Andre, P., \& Montmerle, T.\ 1994, ApJ, 420, 837 

\bibitem[Andre et al.(2000)]{andre00} 
Andre, P., Ward-Thompson, D., \& Barsony, M.\ 2000, Protostars and Planets IV, 59 

\bibitem[Andr{\'e} et al.(2010)]{andre10} 
Andr{\'e}, P., Men'shchikov, A., Bontemps, S., et al.\ 2010, A\&A, 518, L102 

\bibitem[Arce \& Goodman(2002)]{arce02} 
Arce, H.~G., \& Goodman, A.~A.\ 2002, ApJ, 575, 928 

\bibitem[Arce \& Sargent(2006)]{arce06} 
Arce, H.~G., \& Sargent, A.~I.\ 2006, ApJ, 646, 1070 

\bibitem[\protect\citeauthoryear{Arce \etal}{2007}]{arce07} 
Arce, H.~G., Shepherd, D., Gueth, F., Lee, C.-F., Bachiller, R., Rosen, A., \& Beuther, H.\ 2007, Protostars and Planets V, 245 

\bibitem[Bachiller \& Gomez-Gonzalez(1992)]{bachiller92} 
Bachiller, R., \& Gomez-Gonzalez, J.\ 1992, A\&AR, 3, 257

\bibitem[\protect\citeauthoryear{Bate}{1998}]{bate98} 
 Bate, M.~R.\ 1998, ApJL, 508, L95

\bibitem[\protect\citeauthoryear{Bate}{2010}]{bate10} 
 Bate, M.~R.\ 2010, MNRAS, L38

\bibitem[\protect\citeauthoryear{Bate}{2011}]{bate11} 
 Bate, M.~R.\ 2011,  MNRAS, 417, 2036 

\bibitem[\protect\citeauthoryear{Banerjee \& Pudritz}{2006}]{banerjee06}
 Banerjee, R., \& Pudritz, R. E. 2006, ApJ, 641, 949

\bibitem[\protect\citeauthoryear{Bontemps et al.}{1996}]{bontemps96} 
Bontemps, S., Andre, P., Terebey, S., \& Cabrit, S.\ 1996, A\&A, 311, 858 

\bibitem[\protect\citeauthoryear{B{\"u}rzle et al.}{2011}]{burzel11} 
B{\"u}rzle, F., Clark, P.~C., Stasyszyn, F., Dolag, K., \& Klessen, R.~S.\ 2011,  MNRAS, 417, L61 

\bibitem[\protect\citeauthoryear{Blandford \& Payne}{1982}]{blandford82} 
Blandford, R.~D., \& Payne, D.~G.\ 1982, MNRAS, 199, 883 

\bibitem[Cabrit \& Bertout(1992)]{cabrit92} 
Cabrit, S., \& Bertout, C.\ 1992, A\&A, 261, 274 

\bibitem[Cabrit et al.(1997)]{cabrit97} 
Cabrit, S., Raga, A., \& Gueth, F.\ 1997, Herbig-Haro Flows and the Birth of Stars, 182, 163 

\bibitem[Caselli(2002)]{caselli02}
 Caselli, P., Benson, P. J., Myers, P. C., \& Tafalla, M. 2002, ApJ, 572, 238 

\bibitem[Canto \& Raga(1991)]{cant91} 
Canto, J., \& Raga, A.~C.\ 1991, ApJ, 372, 646 

\bibitem[Chen et al.(2010)]{chen10} 
Chen, X., Arce, H.~G., Zhang, Q., et al.\ 2010, ApJ, 715, 1344 

\bibitem[Chen et al.(2012)]{chen12} 
Chen, X., Arce, H.~G., Dunham, M.~M., et al.\ 2012, ApJ, 751, 89 

\bibitem[\protect\citeauthoryear{Commer{\c c}on et al.}{2010}]{commerson10} 
Commer{\c c}on, B., Hennebelle, P., Audit, E., Chabrier, G., \& Teyssier, R.\ 2010, A\&A, 510, L3 

\bibitem[\protect\citeauthoryear{Curtis et al.}{2010}]{curtis10} 
Curtis, E.~I., Richer, J.~S., Swift, J.~J., \& Williams, J.~P.\ 2010, MNRAS, 408, 1516 

\bibitem[\protect\citeauthoryear{Crutcher}{1999}]{crutcher99}
 Crutcher R. M. 1999, ApJ, 520, 706

\bibitem[Downes \& Cabrit(2003)]{downes03} 
Downes, T.~P., \& Cabrit, S.\ 2003, A\&A, 403, 135 

\bibitem[\protect\citeauthoryear{Duffin \& Pudritz}{2009}]{duffin09} 
 Duffin, D.~F., \& Pudritz, R.~E.\ 2009, ApJL, 706, L46 

\bibitem[Duffin et al.(2011)]{duffin11} 
Duffin, D.~F., Pudritz, R.~E., Seifried, D., Banerjee, R., \& Klessen, R.~S.\ 2011, arXiv:1111.5375 

\bibitem[Dunham et al.(2011)]{dunham11} 
Dunham, M.~M., Chen, X., Arce, H.~G., et al.\ 2011, ApJ, 742, 1

\bibitem[Enoch et al.(2009)]{enoch09} 
Enoch, M.~L., Evans, N.~J., II, Sargent, A.~I., \& Glenn, J.\ 2009, ApJ, 692, 973 

\bibitem[Enoch et al.(2010)]{enoch10} 
Enoch, M.~L., Lee, J.-E., Harvey, P., Dunham, M.~M., \& Schnee, S.\ 2010, ApJL, 722, L33 

\bibitem[Evans et al.(2009)]{evans09} 
Evans, N.~J., II, Dunham, M.~M., J{\o}rgensen, J.~K., et al.\ 2009, ApJS, 181, 321 

\bibitem[Goodman \etal(1993)]{goodman93}
 Goodman, A. A., Benson, P. J., Fuller, G. A., \& Myers, P. C. 1993, ApJ, 406, 528

\bibitem[Hartmann et al.(1997)]{HCK97} 
Hartmann, L., Cassen, P., \& Kenyon, S.~J.\ 1997, ApJ, 475, 770

\bibitem[Hatchell et al.(2007)]{hatchell07} 
Hatchell, J., Fuller, G.~A., \& Richer, J.~S.\ 2007, A\&A, 472, 187 

\bibitem[Hennebelle \& Fromang(2008)]{hennebelle08a} 
Hennebelle, P., \& Fromang, S.\ 2008, A\&A, 477, 9 

\bibitem[Hirano et al.(2006)]{hirano06} 
Hirano, N., Liu, S.-Y., Shang, H., et al.\ 2006, ApJL, 636, L141 

\bibitem[Hosokawa \& Omukai(2009)]{HO09} 
Hosokawa, T., \& Omukai, K.\ 2009, ApJ, 691, 823

\bibitem[Hosokawa et al.(2010)]{HYO10} 
Hosokawa, T., Yorke, H.~W., \& Omukai, K.\ 2010, ApJ, 721, 478 

\bibitem[Hosokawa et al.(2011)]{HOK11} 
Hosokawa, T., Offner, S.~S.~R., \& Krumholz, M.~R.\ 2011, ApJ, 738, 140

\bibitem[\protect\citeauthoryear{Inutsuka et al.}{2010}]{inutsuka10} 
 Inutsuka, S., Machida, M.~N., \& Matsumoto, T.\ 2010, ApJL, 718, L58 

\bibitem[\protect\citeauthoryear{Inutsuka}{2012}]{inutsuka12} 
 Inutsuka, 2012, Prog. Theor. Exp. Phys. 01A307 


\bibitem[Kenyon et al.(1990)]{kenyon90} 
Kenyon, S.~J., Hartmann, L.~W., Strom, K.~M., \& Strom, S.~E.\ 1990, AJ, 99, 869 

\bibitem[Kratter et al.(2010)]{kratter10} 
Kratter, K.~M., Matzner, C.~D., Krumholz, M.~R., \& Klein, R.~I.\ 2010, ApJ, 708, 1585 

\bibitem[Kroupa(2001)]{kroupa01} 
Kroupa, P.\ 2001, MNRAS, 322, 231 

\bibitem[Krumholz(2006)]{krumholz06} 
Krumholz, M.~R.\ 2006, ApJL, 641, L45 

\bibitem[\protect\citeauthoryear{Larson}{1969}]{larson69} 
 Larson, R. B., 1969, MNRAS, 145, 271.

\bibitem[Lee et al.(2000)]{lee00} 
Lee, C.-F., Mundy, L.~G., Reipurth, B., Ostriker, E.~C., \& Stone, J.~M.\ 2000, ApJ, 542, 925 

\bibitem[Lizano \& Giovanardi(1995)]{lizano95} 
Lizano, S., \& Giovanardi, C.\ 1995, ApJ, 447, 742 

\bibitem[\protect\citeauthoryear{Machida et al.}{2004}]{machida04} 
 Machida, M. N., Tomisaka, K., \& Matsumoto, T.\ 2004, MNRAS, 348, L1 

\bibitem[\protect\citeauthoryear{Machida \etal}{2005a}]{machida05a}
 Machida, M. N., Matsumoto, T., Tomisaka, K., \& Hanawa, T. 2005, MNRAS, 362, 369

\bibitem[\protect\citeauthoryear{Machida \etal}{2005b}]{machida05b} 
 Machida, M. N., Matsumoto, T., Hanawa, T., \& Tomisaka, K. 2005b, MNRAS, 362, 382 

\bibitem[\protect\citeauthoryear{Machida et al.}{2006}]{machida06} 
 Machida, M.~N., Inutsuka, S., \& Matsumoto, T.\ 2006, ApJL, 647, L151 

\bibitem[\protect\citeauthoryear{Machida et al.}{2007}]{machida07} 
 Machida, M.~N., Inutsuka, S., \& Matsumoto, T.\ 2007, ApJ, 670, 1198 

\bibitem[\protect\citeauthoryear{Machida et al.}{2008b}]{machida08b} 
 Machida, M.~N., Inutsuka, S., \& Matsumoto, T.\ 2008b, ApJ, 676, 1088

\bibitem[\protect\citeauthoryear{Machida et al.}{2010a}]{machida10a} 
 Machida, M.~N., Inutsuka, S., \& Matsumoto, T.\ 2010a, ApJ, 724, 1006 

\bibitem[\protect\citeauthoryear{Machida et al.}{2011a}]{machida11a} 
 Machida, M.~N., Inutsuka, S., \& Matsumoto, T.\ 2011a, ApJ, 729, 42 

\bibitem[Machida et al.(2011b)]{machida11b} 
Machida, M.~N., Inutsuka, S.-I., \& Matsumoto, T.\ 2011b, PASJ, 63, 555 

\bibitem[Machida \& Matsumoto(2011c)]{machida11c} 
Machida, M.~N., \& Matsumoto, T.\ 2011c, MNRAS, 413, 2767 

\bibitem[Machida(2011d)]{machida11d} 
Machida, M.~N.\ 2011d, Computational Star Formation, 270, 65 

\bibitem[Machida \& Matsumoto(2012)]{machida12} 
 Machida, M.~N., \& Matsumoto, T.\ 2012, MNRAS, 421, 588 

\bibitem[\protect\citeauthoryear{Matsumoto \& Tomisaka}{2004}]{matsu04}
 Matsumoto, T., \& Tomisaka, K.\ 2004, ApJ, 616, 266 

\bibitem[\protect\citeauthoryear{Masunaga \& Inutsuka}{2000}]{masunaga00} 
 Masunaga, H., \& Inutsuka, S., 2000, ApJ, 531, 350

\bibitem[Matzner \& McKee(1999)]{matzner99} 
Matzner, C.~D., \& McKee, C.~F.\ 1999, ApJL, 526, L109 

\bibitem[\protect\citeauthoryear{Matzner \& McKee}{2000}]{matzner00} 
 Matzner, C.~D., \& McKee, C.~F.\ 2000, ApJ, 545, 364 

\bibitem[Maury et al.(2011)]{maury11} 
Maury, A.~J., Andr{\'e}, P., Men'shchikov, A., K{\"o}nyves, V., \& Bontemps, S.\ 2011, A\&A, 535, A77 


\bibitem[Mitchell et al.(1994)]{mitchell94} 
Mitchell, G.~F., Hasegawa, T.~I., Dent, W.~R.~F., \& Matthews, H.~E.\ 1994, ApJL, 436, L177 

\bibitem[\protect\citeauthoryear{Mouschovias \& Spitzer}{1976}]{mouschovias76}
 Mouschovias, T. Ch., \& Spitzer, L. 1976, ApJ, 210, 326

\bibitem[Motogi et al.(2011)]{motogi11} 
Motogi, K., Sorai, K., Habe, A., et al.\ 2011, PASJ, 63, 31 

\bibitem[Mundt \& Fried(1983)]{mundt83} 
Mundt, R., \& Fried, J.~W.\ 1983, ApJL, 274, L83 

\bibitem[Nakano et al.(1995)]{nakano95} 
Nakano, T., Hasegawa, T., \& Norman, C.\ 1995, ApJ, 450, 183 

\bibitem[\protect\citeauthoryear{Nakano et al.}{2002}]{nakano02} 
 Nakano, T., Nishi, R., \& Umebayashi, T.\ 2002, ApJ, 573, 199 

\bibitem[Offner \& McKee(2011)]{offner11} 
Offner, S.~S.~R., \& McKee, C.~F.\ 2011, ApJ, 736, 53 

\bibitem[Palla \& Stahler(1991)]{PS91} 
Palla, F., \& Stahler, S.~W.\ 1991, ApJ, 375, 288 

\bibitem[Pineda et al.(2011)]{pineda11} 
Pineda, J.~E., Arce, H.~G., Schnee, S., et al.\ 2011, ApJ, 743, 201 

\bibitem[Price et al.(2012)]{price12} 
Price, D.~J., Tricco, T.~S., \& Bate, M.~R.\ 2012, MNRAS, 423, L45 

\bibitem[Raga \& Cabrit(1993a)]{raga93a} 
Raga, A., \& Cabrit, S.\ 1993a, A\&A, 278, 267 

\bibitem[Raga et al.(1993b)]{raga93b} 
Raga, A.~C., Canto, J., Calvet, N., Rodriguez, L.~F., \& Torrelles, J.~M.\ 1993b, A\&A, 276, 539 

\bibitem[Richer et al.(1992)]{richer92} 
Richer, J.~S., Hills, R.~E., \& Padman, R.\ 1992, MNRAS, 254, 525 

\bibitem[Richer et al.(2000)]{richer00} 
Richer, J.~S., Shepherd, D.~S., Cabrit, S., Bachiller, R., \& Churchwell, E.\ 2000, Protostars and Planets IV, 867 

\bibitem[\protect\citeauthoryear{Saigo \& Tomisaka}{2006}]{saigo06}
 Saigo, K., \& Tomisaka, K.\ 2006, ApJ, 645, 381 

\bibitem[Seifried et al.(2012)]{seifried12} 
Seifried, D., Pudritz, R.~E., Banerjee, R., Duffin, D., \& Klessen, R.~S.\ 2012, MNRAS, 422, 347 

\bibitem[Scott \& Black(1980)]{scott80} 
 Scott, E.~H., \& Black, D.~C.\ 1980, ApJ, 239, 166 

\bibitem[Shu et al.(1991)]{shu91} 
Shu, F.~H., Ruden, S.~P., Lada, C.~J., \& Lizano, S.\ 1991, ApJL, 370, L31 

\bibitem[Snell et al.(1980)]{snell80} 
Snell, R.~L., Loren, R.~B., \& Plambeck, R.~L.\ 1980, ApJL, 239, L17 

\bibitem[Stahler et al.(1980)]{SST80} 
Stahler, S.~W., Shu, F.~H., \& Taam, R.~E.\ 1980, ApJ, 241, 637

\bibitem[Stahler et al.(1980)]{SST80b} 
Stahler, S.~W., Shu, F.~H., \& Taam, R.~E.\ 1980, ApJ, 242, 226 

\bibitem[Stojimirovi{\'c} et al.(2006)]{stoji06} 
Stojimirovi{\'c}, I., Narayanan, G., Snell, R.~L., \& Bally, J.\ 2006, ApJ, 649, 280 

\bibitem[Takahashi \& Ho(2012)]{takahashi12a} 
Takahashi, S., \& Ho, P.~T.~P.\ 2012, ApJL, 745, L10 

\bibitem[Takahashi et al.(2012)]{takahashi12b} 
Takahashi, S., Saigo, K., Ho, P.~T.~P., \& Tomida, K.\ 2012b, ApJ, 752, 10 

\bibitem[Tobin et al.(2012)]{tobin12} 
Tobin, J.~J., Hartmann, L., Chiang, H.-F., et al.\ 2012, Nature, 492, 83 

\bibitem[\protect\citeauthoryear{Tomida et al.}{2010a}]{tomida10a} 
 Tomida, K., Tomisaka, K., Matsumoto, T., Ohsuga, K., Machida, M.~N., \& Saigo, K.\ 2010a, ApJL, 714, L58 

\bibitem[\protect\citeauthoryear{Tomida et al.}{2010b}]{tomida10b} 
Tomida, K., Machida, M.~N., Saigo, K., Tomisaka, K., \& Matsumoto, T.\ 2010b, ApJL, 725, L239 

\bibitem[Tomida et al.(2012)]{tomida12} 
Tomida, K., Tomisaka, K., Matsumoto, T., et al.\ 2012, arXiv:1206.3567 

\bibitem[\protect\citeauthoryear{Tomisaka \etal}{1988a}]{tomisaka88a}
 Tomisaka, K., Ikeuchi, S., \& Nakamura, T. 1988a, ApJ, 326, 208

\bibitem[\protect\citeauthoryear{Tomisaka \etal}{1988b}]{tomisaka88b}
 Tomisaka, K., Ikeuchi, S., \& Nakamura, T. 1988b, ApJ, 335, 239 

\bibitem[Tomisaka(1998)]{tomisaka98} 
Tomisaka, K.\ 1998, ApJL, 502, L163 

\bibitem[Tomisaka(2000)]{tomisaka00} 
Tomisaka, K.\ 2000, ApJL, 528, L41 

\bibitem[\protect\citeauthoryear{Tomisaka}{2002}]{tomisaka02} 
 Tomisaka K., 2002, ApJ, 575, 306

\bibitem[\protect\citeauthoryear{Truelove \etal}{1997}]{truelove97}
 Truelove J, K., Klein R. I., McKee C. F., Holliman J. H., Howell L. H., \& Greenough J. A., 1997, ApJ, 489, L179

\bibitem[\protect\citeauthoryear{Tsukamoto \& Machida}{2011}]{tsukamoto11} 
 Tsukamoto, Y., \& Machida, M.~N.\ 2011, MNRAS, 1146 

\bibitem[Uchida \& Shibata(1985)]{uchida85} 
Uchida, Y., \& Shibata, K.\ 1985, PASJ, 37, 515 

\bibitem[Velusamy \& Langer(1998)]{velusamy98} 
Velusamy, T., \& Langer, W.~D.\ 1998, Nature, 392, 685 

\bibitem[Velusamy et al.(2007)]{velusamy07} 
Velusamy, T., Langer, W.~D., \& Marsh, K.~A.\ 2007, ApJL, 668, L159 

\bibitem[\protect\citeauthoryear{Vorobyov \& Basu}{2006}]{vorobyov06} 
Vorobyov, E.~I., \& Basu, S.\ 2006, ApJ, 650, 956 

\bibitem[\protect\citeauthoryear{Walch et al.}{2009a}]{walch09a} 
Walch, S., Burkert, A., Whitworth, A., Naab, T., \& Gritschneder, M.\ 2009a, MNRAS, 400, 13 


\bibitem[Whitehouse \& Bate(2006)]{whitehouse06} 
Whitehouse, S.~C., \& Bate, M.~R.\ 2006, MNRAS, 367, 32 

\bibitem[\protect\citeauthoryear{Wu \etal}{2004}]{wu04}
 Wu, Y., Wei, Y., Zhao, M., Shi, Y., Yu, W., Qin, S., \& Huang, M. 2004, A\&A, 426, 503

\bibitem[Zhang et al.(2005)]{zhang05} 
Zhang, Q., Hunter, T.~R., Brand, J., et al.\ 2005, ApJ, 625, 864 




\end{thebibliography}
\end{document}